\documentclass[aps,
prd,
twocolumn,
superscriptaddress,
showpacs,
preprintnumbers,
nofootinbib,
floatfix]{revtex4-1}

\usepackage{hyperref}
\usepackage{graphicx}
\usepackage{color}
\usepackage{mathtools}
\usepackage{bm}
\usepackage{braket}
\usepackage{enumerate}
\usepackage[english]{babel}
\usepackage[utf8]{inputenc}

\renewcommand{\k}{\bm{k}}
\newcommand{\p}{\bm{p}}

\newcommand{\Cc}{\mathcal{C}}
\newcommand{\Dc}{\mathcal{D}}

\newcommand{\Kc}{\mathcal{K}}
\newcommand{\Lc}{\mathcal{L}}
\newcommand{\Mc}{\mathcal{M}}

\newcommand{\Oc}{\mathcal{O}}

\DeclareMathOperator{\re}{Re}
\DeclareMathOperator{\im}{Im}

\newcommand{\bcol}{\left[ \begin{array}{c}}
\newcommand{\ecol}{\end{array} \right]}
\newcommand{\beq}{\begin{eqnarray}}
\newcommand{\eeq}{\end{eqnarray}}

\usepackage{mfirstuc} 
\newcommand{\addReviewer}[2]{
  \expandafter\newcommand\csname #1\endcsname[1]{{\bf \color{#2} \capitalisewords{#1}:\,##1}}
  \expandafter\newcommand\csname #1cor\endcsname[2]{{\color{#2} \capitalisewords{#1}:\,\st{##1}{\bf ##2}}}
  \expandafter\newcommand\csname #1color\endcsname{#2}
}

\usepackage{soul,color}
\definecolor{cardinal}{rgb}{0.77, 0.12, 0.23}
\definecolor{asparagus}{rgb}{0.5, 0.70, 0.40}

\addReviewer{sebastian}{cardinal}
\addReviewer{awj}{asparagus}

\hypersetup{ 
    pdfnewwindow=true,   
    colorlinks=true,     
    linkcolor=blue,      
    citecolor=blue,      
    filecolor=blue,      
    urlcolor=blue        
} 

\begin{document}


\newcommand{\UnivWash}{Physics Department, University of Washington, Seattle, WA 98195-1560, USA}
\newcommand{\jlab}{Thomas Jefferson National Accelerator Facility, 12000 Jefferson Avenue, Newport News, Virginia 23606, USA}
\newcommand{\odu}{Department of Physics, Old Dominion University, Norfolk, Virginia 23529, USA}
\newcommand{\UCB}{Department of Physics, University of California, Berkeley, CA 94720, USA}    
\newcommand{\LBNL}{Nuclear Science Division, Lawrence Berkeley National Laboratory, Berkeley, CA 94720, USA}  
\newcommand{\WM}{Department of Physics, William \& Mary, Williamsburg, Virginia 23187, USA}


\title{Evolution of Efimov States}


\author{Sebastian~M.~Dawid}
\email[email: ]{dawids@uw.edu} 
\affiliation{\UnivWash}

\author{Md Habib E Islam} 
\email[email: ]{m2islam@odu.edu}
\affiliation{\odu}
\affiliation{\jlab}

\author{Ra\'ul A.~Brice\~no} 
\email[email: ]{rbriceno@berkeley.edu}
\affiliation{\UCB}
\affiliation{\LBNL}

\author{Andrew W. Jackura}
\email[email: ]{awjackura@wm.edu}
\affiliation{\UCB}
\affiliation{\LBNL}
\affiliation{\WM}


\begin{abstract}
The Efimov phenomenon manifests itself as an emergent discrete scaling symmetry in the quantum three-body problem. In the unitarity limit, it leads to an infinite tower of three-body bound states with energies forming a geometric sequence. In this work, we study the evolution of these so-called Efimov states using relativistic scattering theory. We identify them as poles of the three-particle $S$ matrix and trace their trajectories in the complex energy plane as they evolve from virtual states through bound states to resonances. We dial the scattering parameters toward the unitarity limit and observe the emergence of the universal scaling of energies and couplings---a behavior known from the non-relativistic case. Interestingly, we find that Efimov resonances follow unusual, cyclic trajectories accumulating at the three-body threshold and then disappear at some values of the two-body scattering length. We propose a partial resolution to this ``missing states" problem.
\end{abstract}

\date{\today}
\maketitle


\emph{Introduction:}~The discovery of the Efimov effect in 1970 revealed the formation of an infinite number of bound states, or trimers, in a system of three non-relativistic bosons~\citep{Efimov:1970zz, Efimov:1973awb}. Assuming they interact via two-body forces characterized by a large scattering length, the three-body binding energies form a geometric series with a quotient $\lambda^2 \approx 515$. The emergence of the phenomenon is closely tied to the scale invariance of the quantum-mechanical $1/r^2$ potential~\citep{Case:1950an, deAlfaro:1976vlx, Niemann:2015dsa, Dawid:2017ahd}, and is the best-known example of the renormalization group limit cycle~\citep{PhysRevD.3.1818, Glazek:2002hq, Bedaque:1998kg, Hammer:2011kg}.

The sequence of trimers becomes infinite in the so-called unitarity limit, i.e., when the two-body scattering length, $a$, is made arbitrarily large, $a \to \infty$. While such behavior has not been observed in nature, several nuclear~\cite{Adhikari:1982zz,  Bedaque:1999ve, PhysRevLett.96.112501, Kievsky:2021ghz} and hadronic systems~\citep{Braaten:2003he, Canham:2009zq, Wilbring:2017fwy, Valderrama:2018azi, Valderrama:2018sap} may serve as proxies due to their large scattering length. Furthermore, Efimov physics is realized experimentally using ultracold atoms submerged in a background magnetic field tuned to introduce a Feschbach resonance and drive the system to the unitarity limit~\citep{Kraemer_2006, zaccanti2009observation, doi:10.1126/science.1182840, PhysRevLett.102.165302, PhysRevLett.103.130404, PhysRevLett.103.163202, PhysRevLett.102.140401, 
PhysRevLett.105.103201, PhysRevLett.105.023201, PhysRevLett.105.103203, PhysRevLett.107.120401, PhysRevLett.108.145305, PhysRevLett.112.190401}. Given the generality of the result, this phenomenon has ignited a rich line of research into universality across different subfields~\citep{PhysRevLett.71.4103,  Delfino:1996zz, PhysRevA.74.063604, Braaten:2004rn, PhysRevA.82.043633, von_Stecher_2010, PhysRevLett.108.263001, PhysRevLett.112.105301, PhysRevLett.114.025301, Naidon:2016dpf,  Greene:2017cik, Deltuva:2021fmj, Rosa:2022vql, Frederico:2023fee, Pricoupenko:2023nak}.

\begin{figure*}[t]
    \centering
    \includegraphics[width=0.99\textwidth]{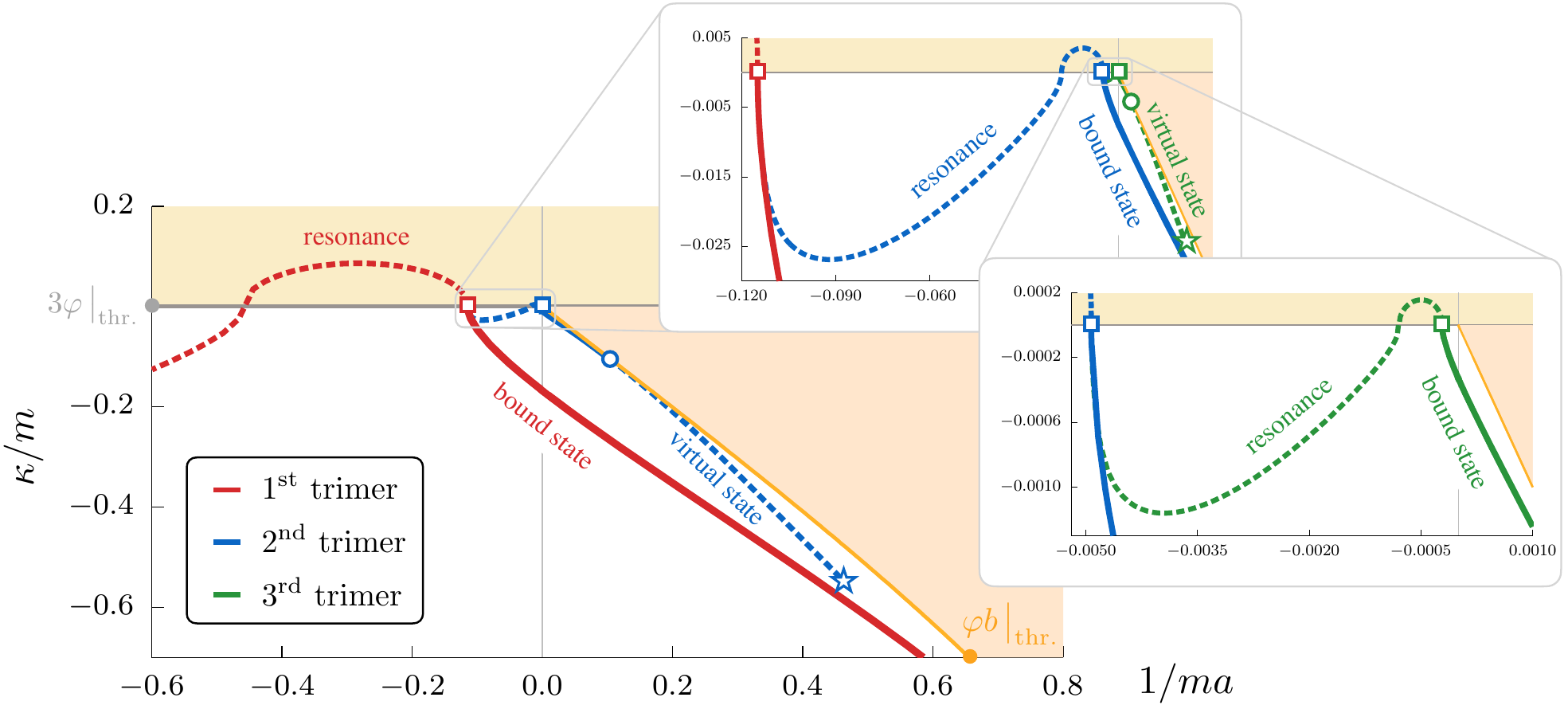}
    \caption{Trajectories of the first three trimer poles in the $(\kappa/m,1/ma)$ plane where $\kappa = {\rm sign} (\re\Delta E)\sqrt{|m \re \Delta E| }$. The $3\varphi$ and $\varphi b$ thresholds are shown explicitly as grey and orange lines. Solid red, blue, and green lines denote physical bound states, while dashed ones denote either virtual bound states on the unphysical $\varphi b$ sheet or resonances on the nearest $ 3\varphi $ sheet. Stars denote the emergence of a virtual state from the logarithmic cut on the second $\varphi b$ sheet. Circles denote the evolution of this virtual state onto a real bound state. Squares denote the further evolution of the state to  three-body resonance. Insets show behavior of trimers near the three-body threshold.}
    \label{fig:kappa_trajectories}
\end{figure*}

Although the unitarity limit does not seem to exist in nature, we can expose the universal scaling behavior by exploring the evolution of Efimov states in vicinity of this limit. We investigate this evolution using relativistic scattering theory, which has been derived as part of ongoing efforts to develop a model-independent framework for studying three-body systems~\cite{Hansen:2014eka, Hansen:2015zga, Mai:2017vot, Blanton:2020gha, Briceno:2017tce, Briceno:2018aml, Hansen:2020zhy, Blanton:2021mih, Blanton:2020jnm, Jackura:2019bmu,  Briceno:2019muc, Blanton:2020gmf, Jackura:2022gib}. Building on previous work~\cite{Dawid:2023jrj}, we identify the trimers as poles of the $S$ matrix in the complex energy variable and study their behavior for various values of $a$, including the $a\to \infty$ limit. We provide evidence of the discrete scaling relationship between the binding energies of the three-body spectrum,
    \beq
    \label{eq:scaling}
    \Delta E_{n}(a) = Q_a^2 \, \Delta E_{n+1}(Q_a a) \, ,
    \eeq
where $\Delta E_n$ is the binding energy of the $n^{\mathrm{th}}$ bound state, and $Q_a$ is a scaling quotient that asymptotes to Efimov's $\lambda$ in the unitarity limit. This scaling relationship holds as the states evolve from bound states to unstable resonances, verifying that the relativistic framework recovers the known non-relativistic results.

Furthermore, by studying the analytic structure of the scattering amplitude, we find a much richer picture of the trimers' behavior than previously identified. We discuss intriguing properties of their evolution across various unphysical Riemann sheets of the complex energy plane, such as the formation of cyclic trajectories of the three-body poles and the emergent scaling property of the associated residues. The behavior of the Efimov resonances is sufficiently puzzling that it motivates us to conjecture about the structure of the three-boson amplitudes and to call for further investigation of these states. Before presenting our findings, we briefly review the framework needed to obtain them.


\emph{Relativistic scattering theory:}~We consider the scattering of three identical spinless bosons of mass $m$, which we label as ``$\varphi$", in their c.m. frame. We fix the total angular momentum of the system to $J=0$, as well as neglect contributions from the two-particle subsystems of angular momenta higher than zero. The $3\varphi \to 3 \varphi$ scattering amplitude depends on the total relativistic energy $E$ and two more variables. We describe the system by splitting the scattering states into a \emph{spectator} particle and a \emph{pair}, formed from the two other bosons, and use initial and final spectator momenta, $k$ and $p$, as the remaining kinematic parameters. In what follows, we use a notation with an implicit energy dependence.

Physical states are associated with poles of the amplitude, with a residue corresponding to the coupling of the state to the open scattering channel. Lehmann–Symanzik–Zimmerman~\citep{Lehmann:1954rq, Zimmermann:1958hg, Zoltan:1960, Duncan:2012aja} reduction implies that this identification also holds for poles off the real energy axis. Causality assures that a complex-valued pole can not reside on the ``physical" energy plane and must instead appear in unphysical Riemann sheets generated by square-root and logarithmic branch cuts of the scattering amplitude. Depending on the location in the complex plane and the sheet, these poles are associated with bound states (real-valued, physical sheet), virtual states (real-valued, unphysical sheet), or resonances (complex-valued, unphysical sheet).

The relativistic three-body amplitude, $\Mc_3(p,k)$, exhibits poles associated with trimers in the $E^2$ plane. Near the $n^{\text{th}}$ pole, it behaves like
    \begin{align}
    \label{eq:M3pole}
    \Mc_{3}(p, k)
    &= -\frac{\Gamma_n(p) \, \Gamma_n(k)}{E^2-E_{n}^2} + \Oc\!\left(E^0\right) \, ,
    \end{align}
where $E_{n}$ is the trimer energy. Bound or virtual states have $\im E_{n} = 0$, while resonances $\im E_{n} \ne 0$. The residue, i.e., the coupling of the $n^{\text{th}}$ trimer to the $3 \varphi$ state, factorizes into momentum-dependent vertex factors $\Gamma_n(k)$ that are closely related to the Faddeev wave functions in the non-relativistic limit.

As implied by the unitarity of the $S$ matrix, the amplitude is described by a set of integral equations~\cite{Hansen:2015zga, Mai:2017vot, Jackura:2018xnx, Jackura:2022gib}. They depend on two dynamical inputs, the $2\varphi \to 2\varphi$ scattering amplitude, $\Mc_2$, and the three-body $K$ matrix, $\Kc_3$, which describes short-distance dynamics of three particles. Given these two objects, one can solve integral equations to obtain the scattering amplitude~\cite{Hansen:2015zga, Jackura:2020bsk, Dawid:2023jrj}. As argued in the supplemental material, the universal scaling behavior is independent of $\Kc_3$, and we set it to zero in the remainder of this letter. Most of the techniques we use have been developed in Ref.~\cite{Dawid:2023jrj} and references within. We discuss some new details in the supplemental material.

The three-body scattering amplitude is given by,
    \begin{align}
    \label{eq:3body_inteq}
    \Mc_3(p,k) & = -\Mc_2(p) \, G(p,k) \, \Mc_2(k) \nonumber \\
    & \qquad - \Mc_2(p) \int_{k'} G(p,k')\,\Mc_3(k',k) \, ,
    \end{align}
where $G$ is the $S$-wave-projected propagator describing particle exchange between the pairs. It is a kinematic function with a logarithmic branch cut~\citep{Hansen:2015zga, Jackura:2018xnx, Dawid:2023jrj}. Finally, the integral measure is $\int_k \equiv \int_0^{k_{\rm max}} {\rm d} k \, k^2 / (2 \pi)^2 \omega_k$, where $k_{\rm max}$ is the maximum allowed value of the momentum and $\omega_k = \sqrt{m^2 + k^2}$ is the spectator energy. The cutoff momentum is fixed by our choice of $\Kc_3 = 0$. Changes in the regularization lead to a different three-body $K$ matrix, assuring that the resultant amplitude is independent of the cutoff.

For $\Mc_2$, we use the leading-order effective range expansion, 
    \beq
    \label{eq:M2_general}
    \Mc_2(k) = \frac{16 \pi \varepsilon_k}{-1/a - i q_k } \, ,
    \eeq
where $\varepsilon_k = \sqrt{(E - \omega_k)^2 - k^2}$ is the pair's energy in its c.m. frame, and $q_k = \sqrt{\varepsilon_k^2 / 4 - m^2}$ is the relative momentum between the particles in the pair. Due to the square root in the definition of the relative momentum, the amplitude is defined on two Riemann branches in the complex $\varepsilon_k$ variable; the first being the physical sheet with $\im q_k > 0$, while the second (unphysical) sheet corresponding to $\im q_k < 0$.

Regardless of the value of $|ma| \geq 1$, the $\Mc_2$ amplitude has a pole in the $\varepsilon_k$ variable, corresponding to a state with mass $m_b  = 2\sqrt{m^2 - 1/a^2}$. It resides on the real axis below the two-body threshold, $\varepsilon_k < 2m$. If $a>0$, the pole is on the first sheet and is associated with a two-body bound state. Otherwise, it is a virtual state on the second sheet.

\begin{figure}[t]
    \centering
    \includegraphics[width=0.48\textwidth, trim = {0 0 0 0}, clip]{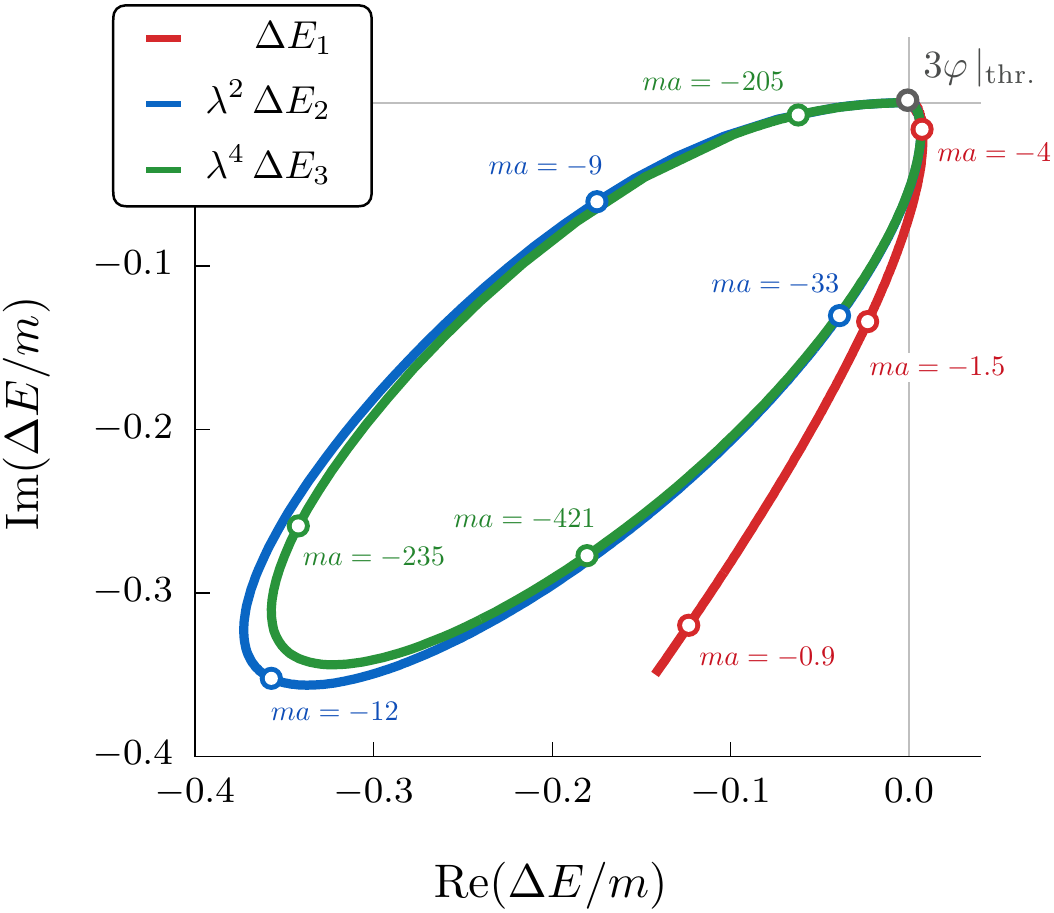}
    \caption{Trajectories of the first three resonances on the nearest unphysical Riemann sheet of the complex $\Delta E$ plane. Energies of the 2nd and 3rd trimers are rescaled by $\lambda^2$ and $\lambda^4$, respectively. At large $|ma|$ and close to the threshold, all trajectories exhibit discrete scaling symmetry. As $|ma|$ decreases, the scaling symmetry breaks down, although, for the second and third resonant states, the discrepancy between the trajectories remains small.}
    \label{fig:3-resonance}
\end{figure}


\emph{Analytic continuation:}~Following our previous work in Refs.~\cite{Jackura:2020bsk, Dawid:2023jrj}, we numerically solve Eq.~\eqref{eq:3body_inteq} to obtain $\Mc_3$ in the complex $E^2$ plane on the physical and the nearest unphysical sheets. Analytic continuation to the complex plane depends on the nature of the singularities of the three-body scattering amplitude encoded in Eq.~\eqref{eq:3body_inteq}.

In addition to potential poles, the $\Mc_3$ amplitude has a logarithmic branch cut inherited from the partial-wave projected propagator, $G$. Furthermore, it has two possible physical thresholds manifesting as corresponding branch points. These are the square-root bound-state-spectator threshold at $E_{\mathrm{thr.}}^{(\varphi b)} = m + m_b$ and the logarithmic three-body threshold at $E_{\mathrm{thr.}}^{(3 \varphi)}=3m$ \citep{LANDAU1959181, Eden:1966dnq}. Unphysical sheets are associated with these two singularities.

The emergence of these thresholds in Eq.~\eqref{eq:3body_inteq} has a non-perturbative origin. The three-body 
amplitude inherits the singularities of $\Mc_2$ in the external momentum variables, $p$ and $k$. The threshold branch points of $\Mc_3(p,k)$ in the $E^2$ plane emerge from the second term of the integral equation when these energy-dependent singularities in the $k'$ variable coincide with the origin of the integration interval, $k'=0$~\citep{Dawid:2023jrj, Eden:1966dnq, burkhardt1969dispersion}. The branch cut at $E=E_{\mathrm{thr.}}^{(\varphi b)}$ arises from the collision of point $k'=0$ with the two-body bound-state pole. The $E_{\mathrm{thr.}}^{(3 \varphi)}$ branch point appears from the collision with the square-root branch point of $\Mc_2$. 

This observation suggests a procedure for the analytic continuation of the amplitude defined by the integral equation. Namely, to extend $\Mc_3$ to the unphysical Riemann sheets of the $E^2$ plane, either through the $\varphi b$ or the $3 \varphi$ cut, one needs to avoid integrating over the discontinuities associated with the above-mentioned collisions, i.e., avoid coincidence of the integration interval with the pole or threshold cut in $k'$. We accomplish this by deforming the integration contour into the complex $k'$ momentum plane in Eq.~\eqref{eq:3body_inteq}. In doing this, we ensure that the deformed integration path avoids logarithmic branch points of $G(p,k')$ and all other singularities induced by the non-perturbative nature of the equation~\cite{Dawid:2017ahd}. We give a more detailed description of this procedure in the supplemental material.

\begin{figure}[t]
    \centering
    \includegraphics[width=0.42\textwidth, trim = {0 0 0 0}, clip]{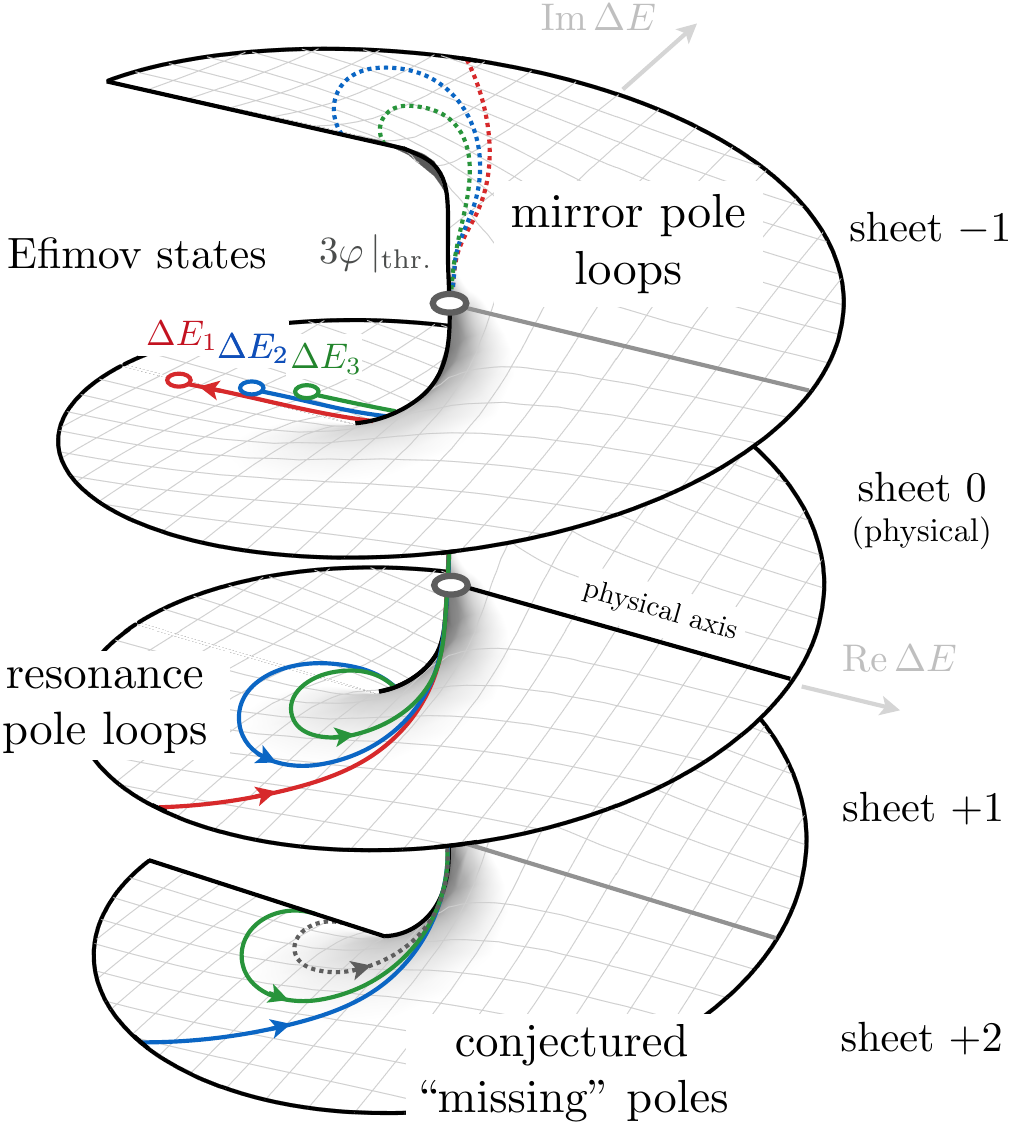}
    \caption{Riemann surfaces of the three-body amplitude in $\Delta E$. The trajectories of the trimers are shown, along with three bound states positions for some $a$. On the nearest unphysical sheet, the 2$^{\rm nd}$ and 3$^{\rm rd}$ trimer exhibit the cyclic behavior as shown in Fig.~\ref{fig:3-resonance}. Mirror poles are found by continuing up to sheet $-1$. We postulate that the higher trimers come from the further unphysical sheets ($\geq +2$), where their cyclic behavior repeats.}
    \label{fig:sheets}
\end{figure}


\emph{Efimov trimers:}~Close to the unitarity limit, i.e., for $|ma| \gg 1$, we find that $\Mc_3$ develops multiple bound state poles. We observe that their binding energies, $\Delta E_n = E_n - E_{\text{thr.}}^{(3\varphi)}$, obey the discrete scaling symmetry given in Eq.~\eqref{eq:scaling}, which is characteristic of the Efimov phenomenon. The quotient $Q_a \to \lambda$ as $ma \to \infty$, confirming that the relativistic framework recovers the expected Efimov scaling. In addition, in the supplement, we show that the vertex functions $\Gamma_n(p)$ agree with the prediction of Ref.~\citep{Hansen:2016ync} for non-relativistic momenta, $p \ll m$. 

By dialing $ma$ to smaller values, we trace the trimers on their trajectories that span across multiple Riemann sheets associated with the dimer-particle and three-particle cuts. The trimers evolve from the virtual states (small, positive $a$) through bound states (large $a$ of both signs) to resonances (small, negative $a$). In Fig.~\ref{fig:kappa_trajectories}, we present their trajectories on the so-called Efimov plot, i.e., in the $(\kappa,1/a)$ plane, where $\kappa = {\rm sign} (\re\Delta E)\sqrt{|m \re \Delta E| }$.

At small values of $a>0$, the ground state exhibits a noticeable deviation from the scaling behavior. Nevertheless, its qualitative features (e.g., pole trajectory, residues) remain analogous to shallow trimers of non-relativistic binding energies.

All excited states follow similar trajectories. They emerge as virtual states on the unphysical $\varphi b$ sheet from the logarithmic cut inherited by $\Mc_3$ from the one-particle exchange amplitude, $G$. They approach the dimer-particle threshold and move to the first sheet, becoming bound states. They remain bound states for large negative values of $a$ and evolve to become resonances on the nearest unphysical sheet associated with the logarithmic $3\varphi$ threshold cut. 

We trace their motion on this sheet, in the complex $\Delta E$ variable, and present it in Fig.~\ref{fig:3-resonance}. At a given finite value of $ma$, there is only a single resonance pole in the unphysical sheet. As we decrease $ma$ from zero to $ma = -8.71$, the ``ground state" resonance moves to the three-body threshold on an arc from complex infinity. It is natural since $ma \to 0$ corresponds to no dynamics and the removal of all but the free states from the spectrum.

By contrast, the excited three-body resonances follow cyclic trajectories, which start and end at the three-body threshold and accumulate near this point. By rescaling them by an appropriate power of $\lambda^2$, we observe that they nearly overlap, providing additional evidence of the discrete scale invariance in the three-boson system. The loop-like trajectories of the Efimov resonances were previously noticed in Ref.~\citep{Hyodo:2013zxa}, where a non-relativistic approach was used.

Moreover, we discover an interesting pattern as these excited trimers move between the physical and unphysical Riemann sheets. Namely, the first excited resonance of energy $\Delta E_2$ emerges from the threshold on the unphysical sheet at the same value, $ma \approx -8.71$, at which the ground state resonance reaches this point and becomes a bound state on the physical energy plane. It leads to a ``missing poles" problem: one pole reaches the threshold, and two emerge. This behavior is repeated for all states, i.e., whenever the $n^{\text{th}}$ resonance enters the threshold, the $(n+1)^{\text{th}}$ resonance appears on the unphysical sheet and the $n^{\text{th}}$ bound state on the physical one. Moreover, we find that the residues of all three poles converge to the same value when they approach the $3 \varphi$ branch point.

This puzzling behavior violates our expectation that the number of poles, equivalent to the number of physical states, must be conserved when one varies the theory parameters. The only exception is the instance of lifted spectrum degeneracy, which we verify does not happen in our system by studying the order of trimer poles.

We propose a possible resolution to this puzzle by noting that the three-body scattering amplitude has infinitely many unphysical sheets; see Fig.~\ref{fig:sheets}. We label the two nearest sheets $\pm 1$ while denoting the physical sheet by $0$. States in the $-1$ plane are the complex-conjugate or ``\emph{mirror}" poles of those in the $+1$ sheet. This is unlike the two-body case, where the Schwarz reflection principle ensures that a resonance pole has its mirror image on the same sheet. For the three-body amplitude, similarly to a complex logarithm, it implies a reflection between the $\pm n$ sheets.

We conjecture that the missing poles come from the higher Riemann branches, one from each. The $n^{\text{th}}$ state approaches the threshold from complex infinity on the $n^{\text{th}}$ Riemann sheet and moves to the $(n-1)^{\text{th}}$ one, where it starts evolving on a cyclic trajectory. Eventually, in the unitarity limit, it travels to the physical energy plane, contributing to the geometric series of bound states. We depict this idea in Fig.~\ref{fig:sheets}, where the dashed lines represent trajectories of $n \geq 4$ states. Reference~\citep{Hyodo:2013zxa} did not address the ``missing poles" issue.

One could verify this conjecture by analytically continuing the amplitude to the higher Riemann sheets. Although we are currently unable to extend our solution to the other sheets, we have performed numerical extrapolations presented in the supplemental material that further support this conjecture. Our proposal only partially resolves the puzzle. Whenever a resonance approaches the threshold, its ``mirror" image does the same. Yet, we observe only one bound state emerging on the physical energy plane. The ``mirror" poles seem to vanish when meeting their complex-conjugate partners at the threshold, which violates our expectation about the conservation of the number of states.


\emph{Discussion:}~To summarize, we found and presented the emergence of the Efimov effect from the relativistic three-body scattering equations. In particular, we discovered evidence of the discrete scaling symmetry in the trajectories of resonances in the nearby unphysical sheet of complex energy. By studying the evolution of the spectrum onto unphysical sheets, we make several observations suggesting that the Efimov phenomenon is closely related to the logarithmic nature of the three-body unitarity cut, i.e., the presence of infinitely many branches. At the same time, our conjecture about the behavior of the trimer trajectories cannot be the end of the story because of the mirror fashion in which $+$ and $-$ sheets contribute trimer poles to the physical sheet.

The ``missing pole" problem is not just a mathematical curiosity but points to a deficiency in our knowledge about the analytic structure and properties of three-particle scattering amplitudes. This, in turn, affects our understanding of the nature of particles that couple strongly to three-particle states~\citep{PhysRevLett.81.5760, COMPASS:2018uzl, PhysRevD.50.4258, Belle:2003nnu, LHCb:2021vvq, LHCb:2021auc, LHCb:2014mir, Cheng:2016shb, Chang:2017wpl, LHCb:2023qne}. Having relativistic scattering amplitudes that satisfy unitarity and whose analytic structure we can fully control will impact a broad set of experimental, phenomenological, and lattice QCD studies. As a result, we close by encouraging further investigations along these lines.


\emph{Acknowledgements:}~The authors thank T. Hyodo, S. Sharpe, M. Baker, and S. Mizera for valuable discussions. SMD is supported by U.S. Department of Energy Contract no.~DE-SC0011637. RAB and MHI acknowledge the support of the USDOE Early Career award, contract DE-SC0019229. MHI acknowledges the support from the Jefferson Science Associates/Jefferson Lab graduate fellowship program. AWJ acknowledges the support of the USDOE ExoHad Topical Collaboration, contract DE-SC0023598.

\bibliographystyle{apsrev4-1}
\bibliography{main}

\begin{thebibliography}{101}%
\makeatletter
\providecommand \@ifxundefined [1]{%
 \@ifx{#1\undefined}
}%
\providecommand \@ifnum [1]{%
 \ifnum #1\expandafter \@firstoftwo
 \else \expandafter \@secondoftwo
 \fi
}%
\providecommand \@ifx [1]{%
 \ifx #1\expandafter \@firstoftwo
 \else \expandafter \@secondoftwo
 \fi
}%
\providecommand \natexlab [1]{#1}%
\providecommand \enquote  [1]{``#1''}%
\providecommand \bibnamefont  [1]{#1}%
\providecommand \bibfnamefont [1]{#1}%
\providecommand \citenamefont [1]{#1}%
\providecommand \href@noop [0]{\@secondoftwo}%
\providecommand \href [0]{\begingroup \@sanitize@url \@href}%
\providecommand \@href[1]{\@@startlink{#1}\@@href}%
\providecommand \@@href[1]{\endgroup#1\@@endlink}%
\providecommand \@sanitize@url [0]{\catcode `\\12\catcode `\$12\catcode
  `\&12\catcode `\#12\catcode `\^12\catcode `\_12\catcode `\%12\relax}%
\providecommand \@@startlink[1]{}%
\providecommand \@@endlink[0]{}%
\providecommand \url  [0]{\begingroup\@sanitize@url \@url }%
\providecommand \@url [1]{\endgroup\@href {#1}{\urlprefix }}%
\providecommand \urlprefix  [0]{URL }%
\providecommand \Eprint [0]{\href }%
\providecommand \doibase [0]{http://dx.doi.org/}%
\providecommand \selectlanguage [0]{\@gobble}%
\providecommand \bibinfo  [0]{\@secondoftwo}%
\providecommand \bibfield  [0]{\@secondoftwo}%
\providecommand \translation [1]{[#1]}%
\providecommand \BibitemOpen [0]{}%
\providecommand \bibitemStop [0]{}%
\providecommand \bibitemNoStop [0]{.\EOS\space}%
\providecommand \EOS [0]{\spacefactor3000\relax}%
\providecommand \BibitemShut  [1]{\csname bibitem#1\endcsname}%
\let\auto@bib@innerbib\@empty
\bibitem [{\citenamefont {Efimov}(1970)}]{Efimov:1970zz}%
  \BibitemOpen
  \bibfield  {author} {\bibinfo {author} {\bibfnamefont {V.}~\bibnamefont
  {Efimov}},\ }\href {\doibase 10.1016/0370-2693(70)90349-7} {\bibfield
  {journal} {\bibinfo  {journal} {Phys. Lett. B}\ }\textbf {\bibinfo {volume}
  {33}},\ \bibinfo {pages} {563} (\bibinfo {year} {1970})}\BibitemShut
  {NoStop}%
\bibitem [{\citenamefont {Efimov}(1973)}]{Efimov:1973awb}%
  \BibitemOpen
  \bibfield  {author} {\bibinfo {author} {\bibfnamefont {V.}~\bibnamefont
  {Efimov}},\ }\href {\doibase 10.1016/0375-9474(73)90510-1} {\bibfield
  {journal} {\bibinfo  {journal} {Nucl. Phys. A}\ }\textbf {\bibinfo {volume}
  {210}},\ \bibinfo {pages} {157} (\bibinfo {year} {1973})}\BibitemShut
  {NoStop}%
\bibitem [{\citenamefont {Case}(1950)}]{Case:1950an}%
  \BibitemOpen
  \bibfield  {author} {\bibinfo {author} {\bibfnamefont {K.~M.}\ \bibnamefont
  {Case}},\ }\href {\doibase 10.1103/PhysRev.80.797} {\bibfield  {journal}
  {\bibinfo  {journal} {Phys. Rev.}\ }\textbf {\bibinfo {volume} {80}},\
  \bibinfo {pages} {797} (\bibinfo {year} {1950})}\BibitemShut {NoStop}%
\bibitem [{\citenamefont {de~Alfaro}\ \emph {et~al.}(1976)\citenamefont
  {de~Alfaro}, \citenamefont {Fubini},\ and\ \citenamefont
  {Furlan}}]{deAlfaro:1976vlx}%
  \BibitemOpen
  \bibfield  {author} {\bibinfo {author} {\bibfnamefont {V.}~\bibnamefont
  {de~Alfaro}}, \bibinfo {author} {\bibfnamefont {S.}~\bibnamefont {Fubini}}, \
  and\ \bibinfo {author} {\bibfnamefont {G.}~\bibnamefont {Furlan}},\ }\href
  {\doibase 10.1007/BF02785666} {\bibfield  {journal} {\bibinfo  {journal}
  {Nuovo Cim. A}\ }\textbf {\bibinfo {volume} {34}},\ \bibinfo {pages} {569}
  (\bibinfo {year} {1976})}\BibitemShut {NoStop}%
\bibitem [{\citenamefont {Niemann}\ and\ \citenamefont
  {Hammer}(2015)}]{Niemann:2015dsa}%
  \BibitemOpen
  \bibfield  {author} {\bibinfo {author} {\bibfnamefont {P.}~\bibnamefont
  {Niemann}}\ and\ \bibinfo {author} {\bibfnamefont {H.~W.}\ \bibnamefont
  {Hammer}},\ }\href {\doibase 10.1007/s00601-015-1001-0} {\bibfield  {journal}
  {\bibinfo  {journal} {Few Body Syst.}\ }\textbf {\bibinfo {volume} {56}},\
  \bibinfo {pages} {869} (\bibinfo {year} {2015})},\ \Eprint
  {http://arxiv.org/abs/1504.04511} {arXiv:1504.04511 [nucl-th]} \BibitemShut
  {NoStop}%
\bibitem [{\citenamefont {Dawid}\ \emph {et~al.}(2018)\citenamefont {Dawid},
  \citenamefont {Gonsior}, \citenamefont {Kwapisz}, \citenamefont {Serafin},
  \citenamefont {Tobolski},\ and\ \citenamefont {G\l{}azek}}]{Dawid:2017ahd}%
  \BibitemOpen
  \bibfield  {author} {\bibinfo {author} {\bibfnamefont {S.~M.}\ \bibnamefont
  {Dawid}}, \bibinfo {author} {\bibfnamefont {R.}~\bibnamefont {Gonsior}},
  \bibinfo {author} {\bibfnamefont {J.}~\bibnamefont {Kwapisz}}, \bibinfo
  {author} {\bibfnamefont {K.}~\bibnamefont {Serafin}}, \bibinfo {author}
  {\bibfnamefont {M.}~\bibnamefont {Tobolski}}, \ and\ \bibinfo {author}
  {\bibfnamefont {S.~D.}\ \bibnamefont {G\l{}azek}},\ }\href {\doibase
  10.1016/j.physletb.2017.12.028} {\bibfield  {journal} {\bibinfo  {journal}
  {Phys. Lett. B}\ }\textbf {\bibinfo {volume} {777}},\ \bibinfo {pages} {260}
  (\bibinfo {year} {2018})},\ \Eprint {http://arxiv.org/abs/1704.08206}
  {arXiv:1704.08206 [quant-ph]} \BibitemShut {NoStop}%
\bibitem [{\citenamefont {Wilson}(1971)}]{PhysRevD.3.1818}%
  \BibitemOpen
  \bibfield  {author} {\bibinfo {author} {\bibfnamefont {K.~G.}\ \bibnamefont
  {Wilson}},\ }\href {\doibase 10.1103/PhysRevD.3.1818} {\bibfield  {journal}
  {\bibinfo  {journal} {Phys. Rev. D}\ }\textbf {\bibinfo {volume} {3}},\
  \bibinfo {pages} {1818} (\bibinfo {year} {1971})}\BibitemShut {NoStop}%
\bibitem [{\citenamefont {Glazek}\ and\ \citenamefont
  {Wilson}(2002)}]{Glazek:2002hq}%
  \BibitemOpen
  \bibfield  {author} {\bibinfo {author} {\bibfnamefont {S.~D.}\ \bibnamefont
  {Glazek}}\ and\ \bibinfo {author} {\bibfnamefont {K.~G.}\ \bibnamefont
  {Wilson}},\ }\href {\doibase 10.1103/PhysRevLett.89.230401} {\bibfield
  {journal} {\bibinfo  {journal} {Phys. Rev. Lett.}\ }\textbf {\bibinfo
  {volume} {89}},\ \bibinfo {pages} {230401} (\bibinfo {year} {2002})},\
  \bibinfo {note} {[Erratum: Phys.Rev.Lett. 92, 139901 (2004)]},\ \Eprint
  {http://arxiv.org/abs/hep-th/0203088} {arXiv:hep-th/0203088} \BibitemShut
  {NoStop}%
\bibitem [{\citenamefont {Bedaque}\ \emph
  {et~al.}(1999{\natexlab{a}})\citenamefont {Bedaque}, \citenamefont {Hammer},\
  and\ \citenamefont {van Kolck}}]{Bedaque:1998kg}%
  \BibitemOpen
  \bibfield  {author} {\bibinfo {author} {\bibfnamefont {P.~F.}\ \bibnamefont
  {Bedaque}}, \bibinfo {author} {\bibfnamefont {H.~W.}\ \bibnamefont {Hammer}},
  \ and\ \bibinfo {author} {\bibfnamefont {U.}~\bibnamefont {van Kolck}},\
  }\href {\doibase 10.1103/PhysRevLett.82.463} {\bibfield  {journal} {\bibinfo
  {journal} {Phys. Rev. Lett.}\ }\textbf {\bibinfo {volume} {82}},\ \bibinfo
  {pages} {463} (\bibinfo {year} {1999}{\natexlab{a}})},\ \Eprint
  {http://arxiv.org/abs/nucl-th/9809025} {arXiv:nucl-th/9809025} \BibitemShut
  {NoStop}%
\bibitem [{\citenamefont {Hammer}\ and\ \citenamefont
  {Platter}(2011)}]{Hammer:2011kg}%
  \BibitemOpen
  \bibfield  {author} {\bibinfo {author} {\bibfnamefont {H.-W.}\ \bibnamefont
  {Hammer}}\ and\ \bibinfo {author} {\bibfnamefont {L.}~\bibnamefont
  {Platter}},\ }\href {\doibase 10.1098/rsta.2011.0001} {\bibfield  {journal}
  {\bibinfo  {journal} {Phil. Trans. Roy. Soc. Lond. A}\ }\textbf {\bibinfo
  {volume} {369}},\ \bibinfo {pages} {2679} (\bibinfo {year} {2011})},\ \Eprint
  {http://arxiv.org/abs/1102.3789} {arXiv:1102.3789 [nucl-th]} \BibitemShut
  {NoStop}%
\bibitem [{\citenamefont {Adhikari}\ and\ \citenamefont
  {Tomio}(1982)}]{Adhikari:1982zz}%
  \BibitemOpen
  \bibfield  {author} {\bibinfo {author} {\bibfnamefont {S.~K.}\ \bibnamefont
  {Adhikari}}\ and\ \bibinfo {author} {\bibfnamefont {L.}~\bibnamefont
  {Tomio}},\ }\href {\doibase 10.1103/PhysRevC.26.83} {\bibfield  {journal}
  {\bibinfo  {journal} {Phys. Rev. C}\ }\textbf {\bibinfo {volume} {26}},\
  \bibinfo {pages} {83} (\bibinfo {year} {1982})}\BibitemShut {NoStop}%
\bibitem [{\citenamefont {Bedaque}\ \emph {et~al.}(2000)\citenamefont
  {Bedaque}, \citenamefont {Hammer},\ and\ \citenamefont {van
  Kolck}}]{Bedaque:1999ve}%
  \BibitemOpen
  \bibfield  {author} {\bibinfo {author} {\bibfnamefont {P.~F.}\ \bibnamefont
  {Bedaque}}, \bibinfo {author} {\bibfnamefont {H.~W.}\ \bibnamefont {Hammer}},
  \ and\ \bibinfo {author} {\bibfnamefont {U.}~\bibnamefont {van Kolck}},\
  }\href {\doibase 10.1016/S0375-9474(00)00205-0} {\bibfield  {journal}
  {\bibinfo  {journal} {Nucl. Phys. A}\ }\textbf {\bibinfo {volume} {676}},\
  \bibinfo {pages} {357} (\bibinfo {year} {2000})},\ \Eprint
  {http://arxiv.org/abs/nucl-th/9906032} {arXiv:nucl-th/9906032} \BibitemShut
  {NoStop}%
\bibitem [{\citenamefont {Garrido}\ \emph {et~al.}(2006)\citenamefont
  {Garrido}, \citenamefont {Fedorov},\ and\ \citenamefont
  {Jensen}}]{PhysRevLett.96.112501}%
  \BibitemOpen
  \bibfield  {author} {\bibinfo {author} {\bibfnamefont {E.}~\bibnamefont
  {Garrido}}, \bibinfo {author} {\bibfnamefont {D.~V.}\ \bibnamefont
  {Fedorov}}, \ and\ \bibinfo {author} {\bibfnamefont {A.~S.}\ \bibnamefont
  {Jensen}},\ }\href {\doibase 10.1103/PhysRevLett.96.112501} {\bibfield
  {journal} {\bibinfo  {journal} {Phys. Rev. Lett.}\ }\textbf {\bibinfo
  {volume} {96}},\ \bibinfo {pages} {112501} (\bibinfo {year}
  {2006})}\BibitemShut {NoStop}%
\bibitem [{\citenamefont {Kievsky}\ \emph {et~al.}(2021)\citenamefont
  {Kievsky}, \citenamefont {Girlanda}, \citenamefont {Gattobigio},\ and\
  \citenamefont {Viviani}}]{Kievsky:2021ghz}%
  \BibitemOpen
  \bibfield  {author} {\bibinfo {author} {\bibfnamefont {A.}~\bibnamefont
  {Kievsky}}, \bibinfo {author} {\bibfnamefont {L.}~\bibnamefont {Girlanda}},
  \bibinfo {author} {\bibfnamefont {M.}~\bibnamefont {Gattobigio}}, \ and\
  \bibinfo {author} {\bibfnamefont {M.}~\bibnamefont {Viviani}},\ }\href
  {\doibase 10.1146/annurev-nucl-102419-032845} {\bibfield  {journal} {\bibinfo
   {journal} {Ann. Rev. Nucl. Part. Sci.}\ }\textbf {\bibinfo {volume} {71}},\
  \bibinfo {pages} {465} (\bibinfo {year} {2021})},\ \Eprint
  {http://arxiv.org/abs/2102.13504} {arXiv:2102.13504 [nucl-th]} \BibitemShut
  {NoStop}%
\bibitem [{\citenamefont {Braaten}\ and\ \citenamefont
  {Kusunoki}(2004)}]{Braaten:2003he}%
  \BibitemOpen
  \bibfield  {author} {\bibinfo {author} {\bibfnamefont {E.}~\bibnamefont
  {Braaten}}\ and\ \bibinfo {author} {\bibfnamefont {M.}~\bibnamefont
  {Kusunoki}},\ }\href {\doibase 10.1103/PhysRevD.69.074005} {\bibfield
  {journal} {\bibinfo  {journal} {Phys. Rev. D}\ }\textbf {\bibinfo {volume}
  {69}},\ \bibinfo {pages} {074005} (\bibinfo {year} {2004})},\ \Eprint
  {http://arxiv.org/abs/hep-ph/0311147} {arXiv:hep-ph/0311147} \BibitemShut
  {NoStop}%
\bibitem [{\citenamefont {Canham}\ \emph {et~al.}(2009)\citenamefont {Canham},
  \citenamefont {Hammer},\ and\ \citenamefont {Springer}}]{Canham:2009zq}%
  \BibitemOpen
  \bibfield  {author} {\bibinfo {author} {\bibfnamefont {D.~L.}\ \bibnamefont
  {Canham}}, \bibinfo {author} {\bibfnamefont {H.~W.}\ \bibnamefont {Hammer}},
  \ and\ \bibinfo {author} {\bibfnamefont {R.~P.}\ \bibnamefont {Springer}},\
  }\href {\doibase 10.1103/PhysRevD.80.014009} {\bibfield  {journal} {\bibinfo
  {journal} {Phys. Rev. D}\ }\textbf {\bibinfo {volume} {80}},\ \bibinfo
  {pages} {014009} (\bibinfo {year} {2009})},\ \Eprint
  {http://arxiv.org/abs/0906.1263} {arXiv:0906.1263 [hep-ph]} \BibitemShut
  {NoStop}%
\bibitem [{\citenamefont {Wilbring}\ \emph {et~al.}(2017)\citenamefont
  {Wilbring}, \citenamefont {Hammer},\ and\ \citenamefont
  {Mei\ss{}ner}}]{Wilbring:2017fwy}%
  \BibitemOpen
  \bibfield  {author} {\bibinfo {author} {\bibfnamefont {E.}~\bibnamefont
  {Wilbring}}, \bibinfo {author} {\bibfnamefont {H.~W.}\ \bibnamefont
  {Hammer}}, \ and\ \bibinfo {author} {\bibfnamefont {U.-G.}\ \bibnamefont
  {Mei\ss{}ner}},\ }\href@noop {} {\  (\bibinfo {year} {2017})},\ \Eprint
  {http://arxiv.org/abs/1705.06176} {arXiv:1705.06176 [hep-ph]} \BibitemShut
  {NoStop}%
\bibitem [{\citenamefont {Valderrama}(2019)}]{Valderrama:2018azi}%
  \BibitemOpen
  \bibfield  {author} {\bibinfo {author} {\bibfnamefont {M.~P.}\ \bibnamefont
  {Valderrama}},\ }\href {\doibase 10.1103/PhysRevD.99.034010} {\bibfield
  {journal} {\bibinfo  {journal} {Phys. Rev. D}\ }\textbf {\bibinfo {volume}
  {99}},\ \bibinfo {pages} {034010} (\bibinfo {year} {2019})},\ \Eprint
  {http://arxiv.org/abs/1811.10173} {arXiv:1811.10173 [hep-ph]} \BibitemShut
  {NoStop}%
\bibitem [{\citenamefont {Valderrama}(2018)}]{Valderrama:2018sap}%
  \BibitemOpen
  \bibfield  {author} {\bibinfo {author} {\bibfnamefont {M.~P.}\ \bibnamefont
  {Valderrama}},\ }\href {\doibase 10.1103/PhysRevD.98.034017} {\bibfield
  {journal} {\bibinfo  {journal} {Phys. Rev. D}\ }\textbf {\bibinfo {volume}
  {98}},\ \bibinfo {pages} {034017} (\bibinfo {year} {2018})},\ \Eprint
  {http://arxiv.org/abs/1805.10584} {arXiv:1805.10584 [hep-ph]} \BibitemShut
  {NoStop}%
\bibitem [{\citenamefont {Kraemer}\ \emph {et~al.}(2006)\citenamefont
  {Kraemer}, \citenamefont {Mark}, \citenamefont {Waldburger}, \citenamefont
  {Danzl}, \citenamefont {Chin}, \citenamefont {Engeser}, \citenamefont
  {Lange}, \citenamefont {Pilch}, \citenamefont {Jaakkola}, \citenamefont
  {Nägerl},\ and\ \citenamefont {Grimm}}]{Kraemer_2006}%
  \BibitemOpen
  \bibfield  {author} {\bibinfo {author} {\bibfnamefont {T.}~\bibnamefont
  {Kraemer}}, \bibinfo {author} {\bibfnamefont {M.}~\bibnamefont {Mark}},
  \bibinfo {author} {\bibfnamefont {P.}~\bibnamefont {Waldburger}}, \bibinfo
  {author} {\bibfnamefont {J.~G.}\ \bibnamefont {Danzl}}, \bibinfo {author}
  {\bibfnamefont {C.}~\bibnamefont {Chin}}, \bibinfo {author} {\bibfnamefont
  {B.}~\bibnamefont {Engeser}}, \bibinfo {author} {\bibfnamefont {A.~D.}\
  \bibnamefont {Lange}}, \bibinfo {author} {\bibfnamefont {K.}~\bibnamefont
  {Pilch}}, \bibinfo {author} {\bibfnamefont {A.}~\bibnamefont {Jaakkola}},
  \bibinfo {author} {\bibfnamefont {H.-C.}\ \bibnamefont {Nägerl}}, \ and\
  \bibinfo {author} {\bibfnamefont {R.}~\bibnamefont {Grimm}},\ }\href
  {\doibase 10.1038/nature04626} {\bibfield  {journal} {\bibinfo  {journal}
  {Nature}\ }\textbf {\bibinfo {volume} {440}},\ \bibinfo {pages} {315}
  (\bibinfo {year} {2006})}\BibitemShut {NoStop}%
\bibitem [{\citenamefont {Zaccanti}\ \emph {et~al.}(2009)\citenamefont
  {Zaccanti}, \citenamefont {Deissler}, \citenamefont {D’Errico},
  \citenamefont {Fattori}, \citenamefont {Jona-Lasinio}, \citenamefont
  {M{\"u}ller}, \citenamefont {Roati}, \citenamefont {Inguscio},\ and\
  \citenamefont {Modugno}}]{zaccanti2009observation}%
  \BibitemOpen
  \bibfield  {author} {\bibinfo {author} {\bibfnamefont {M.}~\bibnamefont
  {Zaccanti}}, \bibinfo {author} {\bibfnamefont {B.}~\bibnamefont {Deissler}},
  \bibinfo {author} {\bibfnamefont {C.}~\bibnamefont {D’Errico}}, \bibinfo
  {author} {\bibfnamefont {M.}~\bibnamefont {Fattori}}, \bibinfo {author}
  {\bibfnamefont {M.}~\bibnamefont {Jona-Lasinio}}, \bibinfo {author}
  {\bibfnamefont {S.}~\bibnamefont {M{\"u}ller}}, \bibinfo {author}
  {\bibfnamefont {G.}~\bibnamefont {Roati}}, \bibinfo {author} {\bibfnamefont
  {M.}~\bibnamefont {Inguscio}}, \ and\ \bibinfo {author} {\bibfnamefont
  {G.}~\bibnamefont {Modugno}},\ }\href@noop {} {\bibfield  {journal} {\bibinfo
   {journal} {Nature Physics}\ }\textbf {\bibinfo {volume} {5}},\ \bibinfo
  {pages} {586} (\bibinfo {year} {2009})}\BibitemShut {NoStop}%
\bibitem [{\citenamefont {Pollack}\ \emph {et~al.}(2009)\citenamefont
  {Pollack}, \citenamefont {Dries},\ and\ \citenamefont
  {Hulet}}]{doi:10.1126/science.1182840}%
  \BibitemOpen
  \bibfield  {author} {\bibinfo {author} {\bibfnamefont {S.~E.}\ \bibnamefont
  {Pollack}}, \bibinfo {author} {\bibfnamefont {D.}~\bibnamefont {Dries}}, \
  and\ \bibinfo {author} {\bibfnamefont {R.~G.}\ \bibnamefont {Hulet}},\ }\href
  {\doibase 10.1126/science.1182840} {\bibfield  {journal} {\bibinfo  {journal}
  {Science}\ }\textbf {\bibinfo {volume} {326}},\ \bibinfo {pages} {1683}
  (\bibinfo {year} {2009})}\BibitemShut {NoStop}%
\bibitem [{\citenamefont {Huckans}\ \emph {et~al.}(2009)\citenamefont
  {Huckans}, \citenamefont {Williams}, \citenamefont {Hazlett}, \citenamefont
  {Stites},\ and\ \citenamefont {O'Hara}}]{PhysRevLett.102.165302}%
  \BibitemOpen
  \bibfield  {author} {\bibinfo {author} {\bibfnamefont {J.~H.}\ \bibnamefont
  {Huckans}}, \bibinfo {author} {\bibfnamefont {J.~R.}\ \bibnamefont
  {Williams}}, \bibinfo {author} {\bibfnamefont {E.~L.}\ \bibnamefont
  {Hazlett}}, \bibinfo {author} {\bibfnamefont {R.~W.}\ \bibnamefont {Stites}},
  \ and\ \bibinfo {author} {\bibfnamefont {K.~M.}\ \bibnamefont {O'Hara}},\
  }\href {\doibase 10.1103/PhysRevLett.102.165302} {\bibfield  {journal}
  {\bibinfo  {journal} {Phys. Rev. Lett.}\ }\textbf {\bibinfo {volume} {102}},\
  \bibinfo {pages} {165302} (\bibinfo {year} {2009})}\BibitemShut {NoStop}%
\bibitem [{\citenamefont {Williams}\ \emph {et~al.}(2009)\citenamefont
  {Williams}, \citenamefont {Hazlett}, \citenamefont {Huckans}, \citenamefont
  {Stites}, \citenamefont {Zhang},\ and\ \citenamefont
  {O'Hara}}]{PhysRevLett.103.130404}%
  \BibitemOpen
  \bibfield  {author} {\bibinfo {author} {\bibfnamefont {J.~R.}\ \bibnamefont
  {Williams}}, \bibinfo {author} {\bibfnamefont {E.~L.}\ \bibnamefont
  {Hazlett}}, \bibinfo {author} {\bibfnamefont {J.~H.}\ \bibnamefont
  {Huckans}}, \bibinfo {author} {\bibfnamefont {R.~W.}\ \bibnamefont {Stites}},
  \bibinfo {author} {\bibfnamefont {Y.}~\bibnamefont {Zhang}}, \ and\ \bibinfo
  {author} {\bibfnamefont {K.~M.}\ \bibnamefont {O'Hara}},\ }\href {\doibase
  10.1103/PhysRevLett.103.130404} {\bibfield  {journal} {\bibinfo  {journal}
  {Phys. Rev. Lett.}\ }\textbf {\bibinfo {volume} {103}},\ \bibinfo {pages}
  {130404} (\bibinfo {year} {2009})}\BibitemShut {NoStop}%
\bibitem [{\citenamefont {Gross}\ \emph {et~al.}(2009)\citenamefont {Gross},
  \citenamefont {Shotan}, \citenamefont {Kokkelmans},\ and\ \citenamefont
  {Khaykovich}}]{PhysRevLett.103.163202}%
  \BibitemOpen
  \bibfield  {author} {\bibinfo {author} {\bibfnamefont {N.}~\bibnamefont
  {Gross}}, \bibinfo {author} {\bibfnamefont {Z.}~\bibnamefont {Shotan}},
  \bibinfo {author} {\bibfnamefont {S.}~\bibnamefont {Kokkelmans}}, \ and\
  \bibinfo {author} {\bibfnamefont {L.}~\bibnamefont {Khaykovich}},\ }\href
  {\doibase 10.1103/PhysRevLett.103.163202} {\bibfield  {journal} {\bibinfo
  {journal} {Phys. Rev. Lett.}\ }\textbf {\bibinfo {volume} {103}},\ \bibinfo
  {pages} {163202} (\bibinfo {year} {2009})}\BibitemShut {NoStop}%
\bibitem [{\citenamefont {Ferlaino}\ \emph {et~al.}(2009)\citenamefont
  {Ferlaino}, \citenamefont {Knoop}, \citenamefont {Berninger}, \citenamefont
  {Harm}, \citenamefont {D'Incao}, \citenamefont {N\"agerl},\ and\
  \citenamefont {Grimm}}]{PhysRevLett.102.140401}%
  \BibitemOpen
  \bibfield  {author} {\bibinfo {author} {\bibfnamefont {F.}~\bibnamefont
  {Ferlaino}}, \bibinfo {author} {\bibfnamefont {S.}~\bibnamefont {Knoop}},
  \bibinfo {author} {\bibfnamefont {M.}~\bibnamefont {Berninger}}, \bibinfo
  {author} {\bibfnamefont {W.}~\bibnamefont {Harm}}, \bibinfo {author}
  {\bibfnamefont {J.~P.}\ \bibnamefont {D'Incao}}, \bibinfo {author}
  {\bibfnamefont {H.-C.}\ \bibnamefont {N\"agerl}}, \ and\ \bibinfo {author}
  {\bibfnamefont {R.}~\bibnamefont {Grimm}},\ }\href {\doibase
  10.1103/PhysRevLett.102.140401} {\bibfield  {journal} {\bibinfo  {journal}
  {Phys. Rev. Lett.}\ }\textbf {\bibinfo {volume} {102}},\ \bibinfo {pages}
  {140401} (\bibinfo {year} {2009})}\BibitemShut {NoStop}%
\bibitem [{\citenamefont {Lompe}\ \emph {et~al.}(2010)\citenamefont {Lompe},
  \citenamefont {Ottenstein}, \citenamefont {Serwane}, \citenamefont {Viering},
  \citenamefont {Wenz}, \citenamefont {Z\"urn},\ and\ \citenamefont
  {Jochim}}]{PhysRevLett.105.103201}%
  \BibitemOpen
  \bibfield  {author} {\bibinfo {author} {\bibfnamefont {T.}~\bibnamefont
  {Lompe}}, \bibinfo {author} {\bibfnamefont {T.~B.}\ \bibnamefont
  {Ottenstein}}, \bibinfo {author} {\bibfnamefont {F.}~\bibnamefont {Serwane}},
  \bibinfo {author} {\bibfnamefont {K.}~\bibnamefont {Viering}}, \bibinfo
  {author} {\bibfnamefont {A.~N.}\ \bibnamefont {Wenz}}, \bibinfo {author}
  {\bibfnamefont {G.}~\bibnamefont {Z\"urn}}, \ and\ \bibinfo {author}
  {\bibfnamefont {S.}~\bibnamefont {Jochim}},\ }\href {\doibase
  10.1103/PhysRevLett.105.103201} {\bibfield  {journal} {\bibinfo  {journal}
  {Phys. Rev. Lett.}\ }\textbf {\bibinfo {volume} {105}},\ \bibinfo {pages}
  {103201} (\bibinfo {year} {2010})}\BibitemShut {NoStop}%
\bibitem [{\citenamefont {Nakajima}\ \emph {et~al.}(2010)\citenamefont
  {Nakajima}, \citenamefont {Horikoshi}, \citenamefont {Mukaiyama},
  \citenamefont {Naidon},\ and\ \citenamefont {Ueda}}]{PhysRevLett.105.023201}%
  \BibitemOpen
  \bibfield  {author} {\bibinfo {author} {\bibfnamefont {S.}~\bibnamefont
  {Nakajima}}, \bibinfo {author} {\bibfnamefont {M.}~\bibnamefont {Horikoshi}},
  \bibinfo {author} {\bibfnamefont {T.}~\bibnamefont {Mukaiyama}}, \bibinfo
  {author} {\bibfnamefont {P.}~\bibnamefont {Naidon}}, \ and\ \bibinfo {author}
  {\bibfnamefont {M.}~\bibnamefont {Ueda}},\ }\href {\doibase
  10.1103/PhysRevLett.105.023201} {\bibfield  {journal} {\bibinfo  {journal}
  {Phys. Rev. Lett.}\ }\textbf {\bibinfo {volume} {105}},\ \bibinfo {pages}
  {023201} (\bibinfo {year} {2010})}\BibitemShut {NoStop}%
\bibitem [{\citenamefont {Gross}\ \emph {et~al.}(2010)\citenamefont {Gross},
  \citenamefont {Shotan}, \citenamefont {Kokkelmans},\ and\ \citenamefont
  {Khaykovich}}]{PhysRevLett.105.103203}%
  \BibitemOpen
  \bibfield  {author} {\bibinfo {author} {\bibfnamefont {N.}~\bibnamefont
  {Gross}}, \bibinfo {author} {\bibfnamefont {Z.}~\bibnamefont {Shotan}},
  \bibinfo {author} {\bibfnamefont {S.}~\bibnamefont {Kokkelmans}}, \ and\
  \bibinfo {author} {\bibfnamefont {L.}~\bibnamefont {Khaykovich}},\ }\href
  {\doibase 10.1103/PhysRevLett.105.103203} {\bibfield  {journal} {\bibinfo
  {journal} {Phys. Rev. Lett.}\ }\textbf {\bibinfo {volume} {105}},\ \bibinfo
  {pages} {103203} (\bibinfo {year} {2010})}\BibitemShut {NoStop}%
\bibitem [{\citenamefont {Berninger}\ \emph {et~al.}(2011)\citenamefont
  {Berninger}, \citenamefont {Zenesini}, \citenamefont {Huang}, \citenamefont
  {Harm}, \citenamefont {N\"agerl}, \citenamefont {Ferlaino}, \citenamefont
  {Grimm}, \citenamefont {Julienne},\ and\ \citenamefont
  {Hutson}}]{PhysRevLett.107.120401}%
  \BibitemOpen
  \bibfield  {author} {\bibinfo {author} {\bibfnamefont {M.}~\bibnamefont
  {Berninger}}, \bibinfo {author} {\bibfnamefont {A.}~\bibnamefont {Zenesini}},
  \bibinfo {author} {\bibfnamefont {B.}~\bibnamefont {Huang}}, \bibinfo
  {author} {\bibfnamefont {W.}~\bibnamefont {Harm}}, \bibinfo {author}
  {\bibfnamefont {H.-C.}\ \bibnamefont {N\"agerl}}, \bibinfo {author}
  {\bibfnamefont {F.}~\bibnamefont {Ferlaino}}, \bibinfo {author}
  {\bibfnamefont {R.}~\bibnamefont {Grimm}}, \bibinfo {author} {\bibfnamefont
  {P.~S.}\ \bibnamefont {Julienne}}, \ and\ \bibinfo {author} {\bibfnamefont
  {J.~M.}\ \bibnamefont {Hutson}},\ }\href {\doibase
  10.1103/PhysRevLett.107.120401} {\bibfield  {journal} {\bibinfo  {journal}
  {Phys. Rev. Lett.}\ }\textbf {\bibinfo {volume} {107}},\ \bibinfo {pages}
  {120401} (\bibinfo {year} {2011})}\BibitemShut {NoStop}%
\bibitem [{\citenamefont {Wild}\ \emph {et~al.}(2012)\citenamefont {Wild},
  \citenamefont {Makotyn}, \citenamefont {Pino}, \citenamefont {Cornell},\ and\
  \citenamefont {Jin}}]{PhysRevLett.108.145305}%
  \BibitemOpen
  \bibfield  {author} {\bibinfo {author} {\bibfnamefont {R.~J.}\ \bibnamefont
  {Wild}}, \bibinfo {author} {\bibfnamefont {P.}~\bibnamefont {Makotyn}},
  \bibinfo {author} {\bibfnamefont {J.~M.}\ \bibnamefont {Pino}}, \bibinfo
  {author} {\bibfnamefont {E.~A.}\ \bibnamefont {Cornell}}, \ and\ \bibinfo
  {author} {\bibfnamefont {D.~S.}\ \bibnamefont {Jin}},\ }\href {\doibase
  10.1103/PhysRevLett.108.145305} {\bibfield  {journal} {\bibinfo  {journal}
  {Phys. Rev. Lett.}\ }\textbf {\bibinfo {volume} {108}},\ \bibinfo {pages}
  {145305} (\bibinfo {year} {2012})}\BibitemShut {NoStop}%
\bibitem [{\citenamefont {Huang}\ \emph {et~al.}(2014)\citenamefont {Huang},
  \citenamefont {Sidorenkov}, \citenamefont {Grimm},\ and\ \citenamefont
  {Hutson}}]{PhysRevLett.112.190401}%
  \BibitemOpen
  \bibfield  {author} {\bibinfo {author} {\bibfnamefont {B.}~\bibnamefont
  {Huang}}, \bibinfo {author} {\bibfnamefont {L.~A.}\ \bibnamefont
  {Sidorenkov}}, \bibinfo {author} {\bibfnamefont {R.}~\bibnamefont {Grimm}}, \
  and\ \bibinfo {author} {\bibfnamefont {J.~M.}\ \bibnamefont {Hutson}},\
  }\href {\doibase 10.1103/PhysRevLett.112.190401} {\bibfield  {journal}
  {\bibinfo  {journal} {Phys. Rev. Lett.}\ }\textbf {\bibinfo {volume} {112}},\
  \bibinfo {pages} {190401} (\bibinfo {year} {2014})}\BibitemShut {NoStop}%
\bibitem [{\citenamefont {Fedorov}\ and\ \citenamefont
  {Jensen}(1993)}]{PhysRevLett.71.4103}%
  \BibitemOpen
  \bibfield  {author} {\bibinfo {author} {\bibfnamefont {D.~V.}\ \bibnamefont
  {Fedorov}}\ and\ \bibinfo {author} {\bibfnamefont {A.~S.}\ \bibnamefont
  {Jensen}},\ }\href {\doibase 10.1103/PhysRevLett.71.4103} {\bibfield
  {journal} {\bibinfo  {journal} {Phys. Rev. Lett.}\ }\textbf {\bibinfo
  {volume} {71}},\ \bibinfo {pages} {4103} (\bibinfo {year}
  {1993})}\BibitemShut {NoStop}%
\bibitem [{\citenamefont {Delfino}\ and\ \citenamefont
  {Frederico}(1996)}]{Delfino:1996zz}%
  \BibitemOpen
  \bibfield  {author} {\bibinfo {author} {\bibfnamefont {A.}~\bibnamefont
  {Delfino}}\ and\ \bibinfo {author} {\bibfnamefont {T.}~\bibnamefont
  {Frederico}},\ }\href {\doibase 10.1103/PhysRevC.53.62} {\bibfield  {journal}
  {\bibinfo  {journal} {Phys. Rev. C}\ }\textbf {\bibinfo {volume} {53}},\
  \bibinfo {pages} {62} (\bibinfo {year} {1996})}\BibitemShut {NoStop}%
\bibitem [{\citenamefont {Hanna}\ and\ \citenamefont
  {Blume}(2006)}]{PhysRevA.74.063604}%
  \BibitemOpen
  \bibfield  {author} {\bibinfo {author} {\bibfnamefont {G.~J.}\ \bibnamefont
  {Hanna}}\ and\ \bibinfo {author} {\bibfnamefont {D.}~\bibnamefont {Blume}},\
  }\href {\doibase 10.1103/PhysRevA.74.063604} {\bibfield  {journal} {\bibinfo
  {journal} {Phys. Rev. A}\ }\textbf {\bibinfo {volume} {74}},\ \bibinfo
  {pages} {063604} (\bibinfo {year} {2006})}\BibitemShut {NoStop}%
\bibitem [{\citenamefont {Braaten}\ and\ \citenamefont
  {Hammer}(2006)}]{Braaten:2004rn}%
  \BibitemOpen
  \bibfield  {author} {\bibinfo {author} {\bibfnamefont {E.}~\bibnamefont
  {Braaten}}\ and\ \bibinfo {author} {\bibfnamefont {H.~W.}\ \bibnamefont
  {Hammer}},\ }\href {\doibase 10.1016/j.physrep.2006.03.001} {\bibfield
  {journal} {\bibinfo  {journal} {Phys. Rept.}\ }\textbf {\bibinfo {volume}
  {428}},\ \bibinfo {pages} {259} (\bibinfo {year} {2006})},\ \Eprint
  {http://arxiv.org/abs/cond-mat/0410417} {arXiv:cond-mat/0410417} \BibitemShut
  {NoStop}%
\bibitem [{\citenamefont {Pricoupenko}(2010)}]{PhysRevA.82.043633}%
  \BibitemOpen
  \bibfield  {author} {\bibinfo {author} {\bibfnamefont {L.}~\bibnamefont
  {Pricoupenko}},\ }\href {\doibase 10.1103/PhysRevA.82.043633} {\bibfield
  {journal} {\bibinfo  {journal} {Phys. Rev. A}\ }\textbf {\bibinfo {volume}
  {82}},\ \bibinfo {pages} {043633} (\bibinfo {year} {2010})}\BibitemShut
  {NoStop}%
\bibitem [{\citenamefont {von Stecher}(2010)}]{von_Stecher_2010}%
  \BibitemOpen
  \bibfield  {author} {\bibinfo {author} {\bibfnamefont {J.}~\bibnamefont {von
  Stecher}},\ }\href {\doibase 10.1088/0953-4075/43/10/101002} {\bibfield
  {journal} {\bibinfo  {journal} {Journal of Physics B: Atomic, Molecular and
  Optical Physics}\ }\textbf {\bibinfo {volume} {43}},\ \bibinfo {pages}
  {101002} (\bibinfo {year} {2010})}\BibitemShut {NoStop}%
\bibitem [{\citenamefont {Wang}\ \emph {et~al.}(2012)\citenamefont {Wang},
  \citenamefont {D'Incao}, \citenamefont {Esry},\ and\ \citenamefont
  {Greene}}]{PhysRevLett.108.263001}%
  \BibitemOpen
  \bibfield  {author} {\bibinfo {author} {\bibfnamefont {J.}~\bibnamefont
  {Wang}}, \bibinfo {author} {\bibfnamefont {J.~P.}\ \bibnamefont {D'Incao}},
  \bibinfo {author} {\bibfnamefont {B.~D.}\ \bibnamefont {Esry}}, \ and\
  \bibinfo {author} {\bibfnamefont {C.~H.}\ \bibnamefont {Greene}},\ }\href
  {\doibase 10.1103/PhysRevLett.108.263001} {\bibfield  {journal} {\bibinfo
  {journal} {Phys. Rev. Lett.}\ }\textbf {\bibinfo {volume} {108}},\ \bibinfo
  {pages} {263001} (\bibinfo {year} {2012})}\BibitemShut {NoStop}%
\bibitem [{\citenamefont {Naidon}\ \emph {et~al.}(2014)\citenamefont {Naidon},
  \citenamefont {Endo},\ and\ \citenamefont {Ueda}}]{PhysRevLett.112.105301}%
  \BibitemOpen
  \bibfield  {author} {\bibinfo {author} {\bibfnamefont {P.}~\bibnamefont
  {Naidon}}, \bibinfo {author} {\bibfnamefont {S.}~\bibnamefont {Endo}}, \ and\
  \bibinfo {author} {\bibfnamefont {M.}~\bibnamefont {Ueda}},\ }\href {\doibase
  10.1103/PhysRevLett.112.105301} {\bibfield  {journal} {\bibinfo  {journal}
  {Phys. Rev. Lett.}\ }\textbf {\bibinfo {volume} {112}},\ \bibinfo {pages}
  {105301} (\bibinfo {year} {2014})}\BibitemShut {NoStop}%
\bibitem [{\citenamefont {Horinouchi}\ and\ \citenamefont
  {Ueda}(2015)}]{PhysRevLett.114.025301}%
  \BibitemOpen
  \bibfield  {author} {\bibinfo {author} {\bibfnamefont {Y.}~\bibnamefont
  {Horinouchi}}\ and\ \bibinfo {author} {\bibfnamefont {M.}~\bibnamefont
  {Ueda}},\ }\href {\doibase 10.1103/PhysRevLett.114.025301} {\bibfield
  {journal} {\bibinfo  {journal} {Phys. Rev. Lett.}\ }\textbf {\bibinfo
  {volume} {114}},\ \bibinfo {pages} {025301} (\bibinfo {year}
  {2015})}\BibitemShut {NoStop}%
\bibitem [{\citenamefont {Naidon}\ and\ \citenamefont
  {Endo}(2017)}]{Naidon:2016dpf}%
  \BibitemOpen
  \bibfield  {author} {\bibinfo {author} {\bibfnamefont {P.}~\bibnamefont
  {Naidon}}\ and\ \bibinfo {author} {\bibfnamefont {S.}~\bibnamefont {Endo}},\
  }\href {\doibase 10.1088/1361-6633/aa50e8} {\bibfield  {journal} {\bibinfo
  {journal} {Rept. Prog. Phys.}\ }\textbf {\bibinfo {volume} {80}},\ \bibinfo
  {pages} {056001} (\bibinfo {year} {2017})},\ \Eprint
  {http://arxiv.org/abs/1610.09805} {arXiv:1610.09805 [quant-ph]} \BibitemShut
  {NoStop}%
\bibitem [{\citenamefont {Greene}\ \emph {et~al.}(2017)\citenamefont {Greene},
  \citenamefont {Giannakeas},\ and\ \citenamefont
  {Perez-Rios}}]{Greene:2017cik}%
  \BibitemOpen
  \bibfield  {author} {\bibinfo {author} {\bibfnamefont {C.~H.}\ \bibnamefont
  {Greene}}, \bibinfo {author} {\bibfnamefont {P.}~\bibnamefont {Giannakeas}},
  \ and\ \bibinfo {author} {\bibfnamefont {J.}~\bibnamefont {Perez-Rios}},\
  }\href {\doibase 10.1103/RevModPhys.89.035006} {\bibfield  {journal}
  {\bibinfo  {journal} {Rev. Mod. Phys.}\ }\textbf {\bibinfo {volume} {89}},\
  \bibinfo {pages} {035006} (\bibinfo {year} {2017})},\ \Eprint
  {http://arxiv.org/abs/1704.02029} {arXiv:1704.02029 [cond-mat.quant-gas]}
  \BibitemShut {NoStop}%
\bibitem [{\citenamefont {Deltuva}(2021)}]{Deltuva:2021fmj}%
  \BibitemOpen
  \bibfield  {author} {\bibinfo {author} {\bibfnamefont {A.}~\bibnamefont
  {Deltuva}},\ }\href {\doibase 10.1103/PhysRevC.103.064001} {\bibfield
  {journal} {\bibinfo  {journal} {Phys. Rev. C}\ }\textbf {\bibinfo {volume}
  {103}},\ \bibinfo {pages} {064001} (\bibinfo {year} {2021})},\ \Eprint
  {http://arxiv.org/abs/2106.01458} {arXiv:2106.01458 [nucl-th]} \BibitemShut
  {NoStop}%
\bibitem [{\citenamefont {Rosa}\ \emph {et~al.}(2022)\citenamefont {Rosa},
  \citenamefont {Frederico}, \citenamefont {Krein},\ and\ \citenamefont
  {Yamashita}}]{Rosa:2022vql}%
  \BibitemOpen
  \bibfield  {author} {\bibinfo {author} {\bibfnamefont {D.~S.}\ \bibnamefont
  {Rosa}}, \bibinfo {author} {\bibfnamefont {T.}~\bibnamefont {Frederico}},
  \bibinfo {author} {\bibfnamefont {G.}~\bibnamefont {Krein}}, \ and\ \bibinfo
  {author} {\bibfnamefont {M.~T.}\ \bibnamefont {Yamashita}},\ }\href {\doibase
  10.1103/PhysRevA.106.023311} {\bibfield  {journal} {\bibinfo  {journal}
  {Phys. Rev. A}\ }\textbf {\bibinfo {volume} {106}},\ \bibinfo {pages}
  {023311} (\bibinfo {year} {2022})},\ \Eprint
  {http://arxiv.org/abs/2209.00575} {arXiv:2209.00575 [physics.atom-ph]}
  \BibitemShut {NoStop}%
\bibitem [{\citenamefont {Frederico}\ and\ \citenamefont
  {Gattobigio}(2023)}]{Frederico:2023fee}%
  \BibitemOpen
  \bibfield  {author} {\bibinfo {author} {\bibfnamefont {T.}~\bibnamefont
  {Frederico}}\ and\ \bibinfo {author} {\bibfnamefont {M.}~\bibnamefont
  {Gattobigio}},\ }\href@noop {} {\  (\bibinfo {year} {2023})},\ \Eprint
  {http://arxiv.org/abs/2303.14952} {arXiv:2303.14952 [physics.atm-clus]}
  \BibitemShut {NoStop}%
\bibitem [{\citenamefont {Pricoupenko}(2023)}]{Pricoupenko:2023nak}%
  \BibitemOpen
  \bibfield  {author} {\bibinfo {author} {\bibfnamefont {L.}~\bibnamefont
  {Pricoupenko}},\ }\href@noop {} {\  (\bibinfo {year} {2023})},\ \Eprint
  {http://arxiv.org/abs/2303.12637} {arXiv:2303.12637 [cond-mat.quant-gas]}
  \BibitemShut {NoStop}%
\bibitem [{\citenamefont {Hansen}\ and\ \citenamefont
  {Sharpe}(2014)}]{Hansen:2014eka}%
  \BibitemOpen
  \bibfield  {author} {\bibinfo {author} {\bibfnamefont {M.~T.}\ \bibnamefont
  {Hansen}}\ and\ \bibinfo {author} {\bibfnamefont {S.~R.}\ \bibnamefont
  {Sharpe}},\ }\href {\doibase 10.1103/PhysRevD.90.116003} {\bibfield
  {journal} {\bibinfo  {journal} {Phys. Rev. D}\ }\textbf {\bibinfo {volume}
  {90}},\ \bibinfo {pages} {116003} (\bibinfo {year} {2014})},\ \Eprint
  {http://arxiv.org/abs/1408.5933} {arXiv:1408.5933 [hep-lat]} \BibitemShut
  {NoStop}%
\bibitem [{\citenamefont {Hansen}\ and\ \citenamefont
  {Sharpe}(2015)}]{Hansen:2015zga}%
  \BibitemOpen
  \bibfield  {author} {\bibinfo {author} {\bibfnamefont {M.~T.}\ \bibnamefont
  {Hansen}}\ and\ \bibinfo {author} {\bibfnamefont {S.~R.}\ \bibnamefont
  {Sharpe}},\ }\href {\doibase 10.1103/PhysRevD.92.114509} {\bibfield
  {journal} {\bibinfo  {journal} {Phys. Rev. D}\ }\textbf {\bibinfo {volume}
  {92}},\ \bibinfo {pages} {114509} (\bibinfo {year} {2015})},\ \Eprint
  {http://arxiv.org/abs/1504.04248} {arXiv:1504.04248 [hep-lat]} \BibitemShut
  {NoStop}%
\bibitem [{\citenamefont {Mai}\ \emph {et~al.}(2017)\citenamefont {Mai},
  \citenamefont {Hu}, \citenamefont {D\"oring}, \citenamefont {Pilloni},\ and\
  \citenamefont {Szczepaniak}}]{Mai:2017vot}%
  \BibitemOpen
  \bibfield  {author} {\bibinfo {author} {\bibfnamefont {M.}~\bibnamefont
  {Mai}}, \bibinfo {author} {\bibfnamefont {B.}~\bibnamefont {Hu}}, \bibinfo
  {author} {\bibfnamefont {M.}~\bibnamefont {D\"oring}}, \bibinfo {author}
  {\bibfnamefont {A.}~\bibnamefont {Pilloni}}, \ and\ \bibinfo {author}
  {\bibfnamefont {A.}~\bibnamefont {Szczepaniak}},\ }\href {\doibase
  10.1140/epja/i2017-12368-4} {\bibfield  {journal} {\bibinfo  {journal}
  {Eur.Phys.J.}\ }\textbf {\bibinfo {volume} {A53}},\ \bibinfo {pages} {177}
  (\bibinfo {year} {2017})},\ \Eprint {http://arxiv.org/abs/1706.06118}
  {arXiv:1706.06118 [nucl-th]} \BibitemShut {NoStop}%
\bibitem [{\citenamefont {Blanton}\ and\ \citenamefont
  {Sharpe}(2020{\natexlab{a}})}]{Blanton:2020gha}%
  \BibitemOpen
  \bibfield  {author} {\bibinfo {author} {\bibfnamefont {T.~D.}\ \bibnamefont
  {Blanton}}\ and\ \bibinfo {author} {\bibfnamefont {S.~R.}\ \bibnamefont
  {Sharpe}},\ }\href {\doibase 10.1103/PhysRevD.102.054520} {\bibfield
  {journal} {\bibinfo  {journal} {Phys. Rev. D}\ }\textbf {\bibinfo {volume}
  {102}},\ \bibinfo {pages} {054520} (\bibinfo {year} {2020}{\natexlab{a}})},\
  \Eprint {http://arxiv.org/abs/2007.16188} {arXiv:2007.16188 [hep-lat]}
  \BibitemShut {NoStop}%
\bibitem [{\citenamefont {Brice\~no}\ \emph {et~al.}(2017)\citenamefont
  {Brice\~no}, \citenamefont {Hansen},\ and\ \citenamefont
  {Sharpe}}]{Briceno:2017tce}%
  \BibitemOpen
  \bibfield  {author} {\bibinfo {author} {\bibfnamefont {R.~A.}\ \bibnamefont
  {Brice\~no}}, \bibinfo {author} {\bibfnamefont {M.~T.}\ \bibnamefont
  {Hansen}}, \ and\ \bibinfo {author} {\bibfnamefont {S.~R.}\ \bibnamefont
  {Sharpe}},\ }\href {\doibase 10.1103/PhysRevD.95.074510} {\bibfield
  {journal} {\bibinfo  {journal} {Phys. Rev. D}\ }\textbf {\bibinfo {volume}
  {95}},\ \bibinfo {pages} {074510} (\bibinfo {year} {2017})},\ \Eprint
  {http://arxiv.org/abs/1701.07465} {arXiv:1701.07465 [hep-lat]} \BibitemShut
  {NoStop}%
\bibitem [{\citenamefont {Brice\~no}\ \emph
  {et~al.}(2019{\natexlab{a}})\citenamefont {Brice\~no}, \citenamefont
  {Hansen},\ and\ \citenamefont {Sharpe}}]{Briceno:2018aml}%
  \BibitemOpen
  \bibfield  {author} {\bibinfo {author} {\bibfnamefont {R.~A.}\ \bibnamefont
  {Brice\~no}}, \bibinfo {author} {\bibfnamefont {M.~T.}\ \bibnamefont
  {Hansen}}, \ and\ \bibinfo {author} {\bibfnamefont {S.~R.}\ \bibnamefont
  {Sharpe}},\ }\href {\doibase 10.1103/PhysRevD.99.014516} {\bibfield
  {journal} {\bibinfo  {journal} {Phys. Rev. D}\ }\textbf {\bibinfo {volume}
  {99}},\ \bibinfo {pages} {014516} (\bibinfo {year} {2019}{\natexlab{a}})},\
  \Eprint {http://arxiv.org/abs/1810.01429} {arXiv:1810.01429 [hep-lat]}
  \BibitemShut {NoStop}%
\bibitem [{\citenamefont {Hansen}\ \emph {et~al.}(2020)\citenamefont {Hansen},
  \citenamefont {Romero-L\'opez},\ and\ \citenamefont
  {Sharpe}}]{Hansen:2020zhy}%
  \BibitemOpen
  \bibfield  {author} {\bibinfo {author} {\bibfnamefont {M.~T.}\ \bibnamefont
  {Hansen}}, \bibinfo {author} {\bibfnamefont {F.}~\bibnamefont
  {Romero-L\'opez}}, \ and\ \bibinfo {author} {\bibfnamefont {S.~R.}\
  \bibnamefont {Sharpe}},\ }\href {\doibase 10.1007/JHEP07(2020)047} {\bibfield
   {journal} {\bibinfo  {journal} {JHEP}\ }\textbf {\bibinfo {volume} {07}},\
  \bibinfo {pages} {047} (\bibinfo {year} {2020})},\ \bibinfo {note} {[Erratum:
  JHEP 02, 014 (2021)]},\ \Eprint {http://arxiv.org/abs/2003.10974}
  {arXiv:2003.10974 [hep-lat]} \BibitemShut {NoStop}%
\bibitem [{\citenamefont {Blanton}\ and\ \citenamefont
  {Sharpe}(2021{\natexlab{a}})}]{Blanton:2021mih}%
  \BibitemOpen
  \bibfield  {author} {\bibinfo {author} {\bibfnamefont {T.~D.}\ \bibnamefont
  {Blanton}}\ and\ \bibinfo {author} {\bibfnamefont {S.~R.}\ \bibnamefont
  {Sharpe}},\ }\href {\doibase 10.1103/PhysRevD.104.034509} {\bibfield
  {journal} {\bibinfo  {journal} {Phys. Rev. D}\ }\textbf {\bibinfo {volume}
  {104}},\ \bibinfo {pages} {034509} (\bibinfo {year} {2021}{\natexlab{a}})},\
  \Eprint {http://arxiv.org/abs/2105.12094} {arXiv:2105.12094 [hep-lat]}
  \BibitemShut {NoStop}%
\bibitem [{\citenamefont {Blanton}\ and\ \citenamefont
  {Sharpe}(2020{\natexlab{b}})}]{Blanton:2020jnm}%
  \BibitemOpen
  \bibfield  {author} {\bibinfo {author} {\bibfnamefont {T.~D.}\ \bibnamefont
  {Blanton}}\ and\ \bibinfo {author} {\bibfnamefont {S.~R.}\ \bibnamefont
  {Sharpe}},\ }\href {\doibase 10.1103/PhysRevD.102.054515} {\bibfield
  {journal} {\bibinfo  {journal} {Phys. Rev. D}\ }\textbf {\bibinfo {volume}
  {102}},\ \bibinfo {pages} {054515} (\bibinfo {year} {2020}{\natexlab{b}})},\
  \Eprint {http://arxiv.org/abs/2007.16190} {arXiv:2007.16190 [hep-lat]}
  \BibitemShut {NoStop}%
\bibitem [{\citenamefont {Jackura}\ \emph
  {et~al.}(2019{\natexlab{a}})\citenamefont {Jackura}, \citenamefont {Dawid},
  \citenamefont {Fern\'andez-Ram\'\i{}rez}, \citenamefont {Mathieu},
  \citenamefont {Mikhasenko}, \citenamefont {Pilloni}, \citenamefont {Sharpe},\
  and\ \citenamefont {Szczepaniak}}]{Jackura:2019bmu}%
  \BibitemOpen
  \bibfield  {author} {\bibinfo {author} {\bibfnamefont {A.~W.}\ \bibnamefont
  {Jackura}}, \bibinfo {author} {\bibfnamefont {S.~M.}\ \bibnamefont {Dawid}},
  \bibinfo {author} {\bibfnamefont {C.}~\bibnamefont
  {Fern\'andez-Ram\'\i{}rez}}, \bibinfo {author} {\bibfnamefont
  {V.}~\bibnamefont {Mathieu}}, \bibinfo {author} {\bibfnamefont
  {M.}~\bibnamefont {Mikhasenko}}, \bibinfo {author} {\bibfnamefont
  {A.}~\bibnamefont {Pilloni}}, \bibinfo {author} {\bibfnamefont {S.~R.}\
  \bibnamefont {Sharpe}}, \ and\ \bibinfo {author} {\bibfnamefont {A.~P.}\
  \bibnamefont {Szczepaniak}},\ }\href {\doibase 10.1103/PhysRevD.100.034508}
  {\bibfield  {journal} {\bibinfo  {journal} {Phys. Rev. D}\ }\textbf {\bibinfo
  {volume} {100}},\ \bibinfo {pages} {034508} (\bibinfo {year}
  {2019}{\natexlab{a}})},\ \Eprint {http://arxiv.org/abs/1905.12007}
  {arXiv:1905.12007 [hep-ph]} \BibitemShut {NoStop}%
\bibitem [{\citenamefont {Brice\~no}\ \emph
  {et~al.}(2019{\natexlab{b}})\citenamefont {Brice\~no}, \citenamefont
  {Hansen}, \citenamefont {Sharpe},\ and\ \citenamefont
  {Szczepaniak}}]{Briceno:2019muc}%
  \BibitemOpen
  \bibfield  {author} {\bibinfo {author} {\bibfnamefont {R.~A.}\ \bibnamefont
  {Brice\~no}}, \bibinfo {author} {\bibfnamefont {M.~T.}\ \bibnamefont
  {Hansen}}, \bibinfo {author} {\bibfnamefont {S.~R.}\ \bibnamefont {Sharpe}},
  \ and\ \bibinfo {author} {\bibfnamefont {A.~P.}\ \bibnamefont
  {Szczepaniak}},\ }\href {\doibase 10.1103/PhysRevD.100.054508} {\bibfield
  {journal} {\bibinfo  {journal} {Phys. Rev. D}\ }\textbf {\bibinfo {volume}
  {100}},\ \bibinfo {pages} {054508} (\bibinfo {year} {2019}{\natexlab{b}})},\
  \Eprint {http://arxiv.org/abs/1905.11188} {arXiv:1905.11188 [hep-lat]}
  \BibitemShut {NoStop}%
\bibitem [{\citenamefont {Blanton}\ and\ \citenamefont
  {Sharpe}(2021{\natexlab{b}})}]{Blanton:2020gmf}%
  \BibitemOpen
  \bibfield  {author} {\bibinfo {author} {\bibfnamefont {T.~D.}\ \bibnamefont
  {Blanton}}\ and\ \bibinfo {author} {\bibfnamefont {S.~R.}\ \bibnamefont
  {Sharpe}},\ }\href {\doibase 10.1103/PhysRevD.103.054503} {\bibfield
  {journal} {\bibinfo  {journal} {Phys. Rev. D}\ }\textbf {\bibinfo {volume}
  {103}},\ \bibinfo {pages} {054503} (\bibinfo {year} {2021}{\natexlab{b}})},\
  \Eprint {http://arxiv.org/abs/2011.05520} {arXiv:2011.05520 [hep-lat]}
  \BibitemShut {NoStop}%
\bibitem [{\citenamefont {Jackura}(2023)}]{Jackura:2022gib}%
  \BibitemOpen
  \bibfield  {author} {\bibinfo {author} {\bibfnamefont {A.~W.}\ \bibnamefont
  {Jackura}},\ }\href {\doibase 10.1103/PhysRevD.108.034505} {\bibfield
  {journal} {\bibinfo  {journal} {Phys. Rev. D}\ }\textbf {\bibinfo {volume}
  {108}},\ \bibinfo {pages} {034505} (\bibinfo {year} {2023})}\BibitemShut
  {NoStop}%
\bibitem [{\citenamefont {Dawid}\ \emph {et~al.}(2023)\citenamefont {Dawid},
  \citenamefont {Islam},\ and\ \citenamefont {Brice\~no}}]{Dawid:2023jrj}%
  \BibitemOpen
  \bibfield  {author} {\bibinfo {author} {\bibfnamefont {S.~M.}\ \bibnamefont
  {Dawid}}, \bibinfo {author} {\bibfnamefont {M.~H.~E.}\ \bibnamefont {Islam}},
  \ and\ \bibinfo {author} {\bibfnamefont {R.~A.}\ \bibnamefont {Brice\~no}},\
  }\href {\doibase 10.1103/PhysRevD.108.034016} {\bibfield  {journal} {\bibinfo
   {journal} {Phys. Rev. D}\ }\textbf {\bibinfo {volume} {108}},\ \bibinfo
  {pages} {034016} (\bibinfo {year} {2023})}\BibitemShut {NoStop}%
\bibitem [{\citenamefont {Lehmann}\ \emph {et~al.}(1955)\citenamefont
  {Lehmann}, \citenamefont {Symanzik},\ and\ \citenamefont
  {Zimmermann}}]{Lehmann:1954rq}%
  \BibitemOpen
  \bibfield  {author} {\bibinfo {author} {\bibfnamefont {H.}~\bibnamefont
  {Lehmann}}, \bibinfo {author} {\bibfnamefont {K.}~\bibnamefont {Symanzik}}, \
  and\ \bibinfo {author} {\bibfnamefont {W.}~\bibnamefont {Zimmermann}},\
  }\href {\doibase 10.1007/BF02731765} {\bibfield  {journal} {\bibinfo
  {journal} {Nuovo Cim.}\ }\textbf {\bibinfo {volume} {1}},\ \bibinfo {pages}
  {205} (\bibinfo {year} {1955})}\BibitemShut {NoStop}%
\bibitem [{\citenamefont {Zimmermann}(1958)}]{Zimmermann:1958hg}%
  \BibitemOpen
  \bibfield  {author} {\bibinfo {author} {\bibfnamefont {W.}~\bibnamefont
  {Zimmermann}},\ }\href {\doibase 10.1007/BF02859796} {\bibfield  {journal}
  {\bibinfo  {journal} {Nuovo Cim.}\ }\textbf {\bibinfo {volume} {10}},\
  \bibinfo {pages} {597} (\bibinfo {year} {1958})}\BibitemShut {NoStop}%
\bibitem [{\citenamefont {Fried}(1960)}]{Zoltan:1960}%
  \BibitemOpen
  \bibfield  {author} {\bibinfo {author} {\bibfnamefont {Z.}~\bibnamefont
  {Fried}},\ }{\selectlanguage {English}\emph {\bibinfo {title} {Bound states
  in the Lehmann, Symanzik and Zimmermann formulation of field theory}}},\
  \href
  {https://www.proquest.com/dissertations-theses/bound-states-lehmann-symanzik-zimmermann/docview/301856096/se-2}
  {Ph.D. thesis} (\bibinfo {year} {1960})\BibitemShut {NoStop}%
\bibitem [{\citenamefont {Duncan}(2012)}]{Duncan:2012aja}%
  \BibitemOpen
  \bibfield  {author} {\bibinfo {author} {\bibfnamefont {A.}~\bibnamefont
  {Duncan}},\ }\href {\doibase 10.1093/acprof:oso/9780199573264.001.0001}
  {\emph {\bibinfo {title} {{The Conceptual Framework of Quantum Field
  Theory}}}}\ (\bibinfo  {publisher} {Oxford University Press},\ \bibinfo
  {year} {2012})\BibitemShut {NoStop}%
\bibitem [{\citenamefont {Jackura}\ \emph
  {et~al.}(2019{\natexlab{b}})\citenamefont {Jackura}, \citenamefont
  {Fern\'andez-Ram\'\i{}rez}, \citenamefont {Mathieu}, \citenamefont
  {Mikhasenko}, \citenamefont {Nys}, \citenamefont {Pilloni}, \citenamefont
  {Salda\~na}, \citenamefont {Sherrill},\ and\ \citenamefont
  {Szczepaniak}}]{Jackura:2018xnx}%
  \BibitemOpen
  \bibfield  {author} {\bibinfo {author} {\bibfnamefont {A.}~\bibnamefont
  {Jackura}}, \bibinfo {author} {\bibfnamefont {C.}~\bibnamefont
  {Fern\'andez-Ram\'\i{}rez}}, \bibinfo {author} {\bibfnamefont
  {V.}~\bibnamefont {Mathieu}}, \bibinfo {author} {\bibfnamefont
  {M.}~\bibnamefont {Mikhasenko}}, \bibinfo {author} {\bibfnamefont
  {J.}~\bibnamefont {Nys}}, \bibinfo {author} {\bibfnamefont {A.}~\bibnamefont
  {Pilloni}}, \bibinfo {author} {\bibfnamefont {K.}~\bibnamefont {Salda\~na}},
  \bibinfo {author} {\bibfnamefont {N.}~\bibnamefont {Sherrill}}, \ and\
  \bibinfo {author} {\bibfnamefont {A.~P.}\ \bibnamefont {Szczepaniak}}
  (\bibinfo {collaboration} {JPAC}),\ }\href {\doibase
  10.1140/epjc/s10052-019-6566-1} {\bibfield  {journal} {\bibinfo  {journal}
  {Eur. Phys. J. C}\ }\textbf {\bibinfo {volume} {79}},\ \bibinfo {pages} {56}
  (\bibinfo {year} {2019}{\natexlab{b}})},\ \Eprint
  {http://arxiv.org/abs/1809.10523} {arXiv:1809.10523 [hep-ph]} \BibitemShut
  {NoStop}%
\bibitem [{\citenamefont {Jackura}\ \emph {et~al.}(2021)\citenamefont
  {Jackura}, \citenamefont {Brice\~no}, \citenamefont {Dawid}, \citenamefont
  {Islam},\ and\ \citenamefont {McCarty}}]{Jackura:2020bsk}%
  \BibitemOpen
  \bibfield  {author} {\bibinfo {author} {\bibfnamefont {A.~W.}\ \bibnamefont
  {Jackura}}, \bibinfo {author} {\bibfnamefont {R.~A.}\ \bibnamefont
  {Brice\~no}}, \bibinfo {author} {\bibfnamefont {S.~M.}\ \bibnamefont
  {Dawid}}, \bibinfo {author} {\bibfnamefont {M.~H.~E.}\ \bibnamefont {Islam}},
  \ and\ \bibinfo {author} {\bibfnamefont {C.}~\bibnamefont {McCarty}},\ }\href
  {\doibase 10.1103/PhysRevD.104.014507} {\bibfield  {journal} {\bibinfo
  {journal} {Phys. Rev. D}\ }\textbf {\bibinfo {volume} {104}},\ \bibinfo
  {pages} {014507} (\bibinfo {year} {2021})},\ \Eprint
  {http://arxiv.org/abs/2010.09820} {arXiv:2010.09820 [hep-lat]} \BibitemShut
  {NoStop}%
\bibitem [{\citenamefont {Landau}(1959)}]{LANDAU1959181}%
  \BibitemOpen
  \bibfield  {author} {\bibinfo {author} {\bibfnamefont {L.}~\bibnamefont
  {Landau}},\ }\href {\doibase https://doi.org/10.1016/0029-5582(59)90154-3}
  {\bibfield  {journal} {\bibinfo  {journal} {Nuclear Physics}\ }\textbf
  {\bibinfo {volume} {13}},\ \bibinfo {pages} {181} (\bibinfo {year}
  {1959})}\BibitemShut {NoStop}%
\bibitem [{\citenamefont {Eden}\ \emph {et~al.}(1966)\citenamefont {Eden},
  \citenamefont {Landshoff}, \citenamefont {Olive},\ and\ \citenamefont
  {Polkinghorne}}]{Eden:1966dnq}%
  \BibitemOpen
  \bibfield  {author} {\bibinfo {author} {\bibfnamefont {R.~J.}\ \bibnamefont
  {Eden}}, \bibinfo {author} {\bibfnamefont {P.~V.}\ \bibnamefont {Landshoff}},
  \bibinfo {author} {\bibfnamefont {D.~I.}\ \bibnamefont {Olive}}, \ and\
  \bibinfo {author} {\bibfnamefont {J.~C.}\ \bibnamefont {Polkinghorne}},\
  }\href@noop {} {\emph {\bibinfo {title} {{The analytic S-matrix}}}}\
  (\bibinfo  {publisher} {Cambridge Univ. Press},\ \bibinfo {address}
  {Cambridge},\ \bibinfo {year} {1966})\BibitemShut {NoStop}%
\bibitem [{\citenamefont {Burkhardt}(1969)}]{burkhardt1969dispersion}%
  \BibitemOpen
  \bibfield  {author} {\bibinfo {author} {\bibfnamefont {H.}~\bibnamefont
  {Burkhardt}},\ }\href {https://books.google.com/books?id=oFReAAAAIAAJ} {\emph
  {\bibinfo {title} {Dispersion Relation Dynamics: A Phenomenological
  Introduction to S-matrix Theory}}}\ (\bibinfo  {publisher} {North-Holland
  Publishing Company},\ \bibinfo {year} {1969})\BibitemShut {NoStop}%
\bibitem [{\citenamefont {Hansen}\ and\ \citenamefont
  {Sharpe}(2017)}]{Hansen:2016ync}%
  \BibitemOpen
  \bibfield  {author} {\bibinfo {author} {\bibfnamefont {M.~T.}\ \bibnamefont
  {Hansen}}\ and\ \bibinfo {author} {\bibfnamefont {S.~R.}\ \bibnamefont
  {Sharpe}},\ }\href {\doibase 10.1103/PhysRevD.95.034501} {\bibfield
  {journal} {\bibinfo  {journal} {Phys. Rev. D}\ }\textbf {\bibinfo {volume}
  {95}},\ \bibinfo {pages} {034501} (\bibinfo {year} {2017})},\ \Eprint
  {http://arxiv.org/abs/1609.04317} {arXiv:1609.04317 [hep-lat]} \BibitemShut
  {NoStop}%
\bibitem [{\citenamefont {Hyodo}\ \emph {et~al.}(2014)\citenamefont {Hyodo},
  \citenamefont {Hatsuda},\ and\ \citenamefont {Nishida}}]{Hyodo:2013zxa}%
  \BibitemOpen
  \bibfield  {author} {\bibinfo {author} {\bibfnamefont {T.}~\bibnamefont
  {Hyodo}}, \bibinfo {author} {\bibfnamefont {T.}~\bibnamefont {Hatsuda}}, \
  and\ \bibinfo {author} {\bibfnamefont {Y.}~\bibnamefont {Nishida}},\ }\href
  {\doibase 10.1103/PhysRevC.89.032201} {\bibfield  {journal} {\bibinfo
  {journal} {Phys. Rev. C}\ }\textbf {\bibinfo {volume} {89}},\ \bibinfo
  {pages} {032201} (\bibinfo {year} {2014})},\ \Eprint
  {http://arxiv.org/abs/1311.6289} {arXiv:1311.6289 [hep-ph]} \BibitemShut
  {NoStop}%
\bibitem [{\citenamefont {Adams}\ \emph {et~al.}(1998)\citenamefont {Adams}
  \emph {et~al.}}]{PhysRevLett.81.5760}%
  \BibitemOpen
  \bibfield  {author} {\bibinfo {author} {\bibfnamefont {G.~S.}\ \bibnamefont
  {Adams}} \emph {et~al.} (\bibinfo {collaboration} {E852 Collaboration}),\
  }\href {\doibase 10.1103/PhysRevLett.81.5760} {\bibfield  {journal} {\bibinfo
   {journal} {Phys. Rev. Lett.}\ }\textbf {\bibinfo {volume} {81}},\ \bibinfo
  {pages} {5760} (\bibinfo {year} {1998})}\BibitemShut {NoStop}%
\bibitem [{\citenamefont {Aghasyan}\ \emph {et~al.}(2018)\citenamefont
  {Aghasyan} \emph {et~al.}}]{COMPASS:2018uzl}%
  \BibitemOpen
  \bibfield  {author} {\bibinfo {author} {\bibfnamefont {M.}~\bibnamefont
  {Aghasyan}} \emph {et~al.} (\bibinfo {collaboration} {COMPASS}),\ }\href
  {\doibase 10.1103/PhysRevD.98.092003} {\bibfield  {journal} {\bibinfo
  {journal} {Phys. Rev. D}\ }\textbf {\bibinfo {volume} {98}},\ \bibinfo
  {pages} {092003} (\bibinfo {year} {2018})},\ \Eprint
  {http://arxiv.org/abs/1802.05913} {arXiv:1802.05913 [hep-ex]} \BibitemShut
  {NoStop}%
\bibitem [{\citenamefont {Antoniazzi}\ \emph {et~al.}(1994)\citenamefont
  {Antoniazzi} \emph {et~al.}}]{PhysRevD.50.4258}%
  \BibitemOpen
  \bibfield  {author} {\bibinfo {author} {\bibfnamefont {L.}~\bibnamefont
  {Antoniazzi}} \emph {et~al.},\ }\href {\doibase 10.1103/PhysRevD.50.4258}
  {\bibfield  {journal} {\bibinfo  {journal} {Phys. Rev. D}\ }\textbf {\bibinfo
  {volume} {50}},\ \bibinfo {pages} {4258} (\bibinfo {year}
  {1994})}\BibitemShut {NoStop}%
\bibitem [{\citenamefont {Choi}\ \emph {et~al.}(2003)\citenamefont {Choi} \emph
  {et~al.}}]{Belle:2003nnu}%
  \BibitemOpen
  \bibfield  {author} {\bibinfo {author} {\bibfnamefont {S.~K.}\ \bibnamefont
  {Choi}} \emph {et~al.} (\bibinfo {collaboration} {Belle}),\ }\href {\doibase
  10.1103/PhysRevLett.91.262001} {\bibfield  {journal} {\bibinfo  {journal}
  {Phys. Rev. Lett.}\ }\textbf {\bibinfo {volume} {91}},\ \bibinfo {pages}
  {262001} (\bibinfo {year} {2003})},\ \Eprint
  {http://arxiv.org/abs/hep-ex/0309032} {arXiv:hep-ex/0309032} \BibitemShut
  {NoStop}%
\bibitem [{\citenamefont {Aaij}\ \emph
  {et~al.}(2022{\natexlab{a}})\citenamefont {Aaij} \emph
  {et~al.}}]{LHCb:2021vvq}%
  \BibitemOpen
  \bibfield  {author} {\bibinfo {author} {\bibfnamefont {R.}~\bibnamefont
  {Aaij}} \emph {et~al.} (\bibinfo {collaboration} {LHCb}),\ }\href {\doibase
  10.1038/s41567-022-01614-y} {\bibfield  {journal} {\bibinfo  {journal}
  {Nature Phys.}\ }\textbf {\bibinfo {volume} {18}},\ \bibinfo {pages} {751}
  (\bibinfo {year} {2022}{\natexlab{a}})},\ \Eprint
  {http://arxiv.org/abs/2109.01038} {arXiv:2109.01038 [hep-ex]} \BibitemShut
  {NoStop}%
\bibitem [{\citenamefont {Aaij}\ \emph
  {et~al.}(2022{\natexlab{b}})\citenamefont {Aaij} \emph
  {et~al.}}]{LHCb:2021auc}%
  \BibitemOpen
  \bibfield  {author} {\bibinfo {author} {\bibfnamefont {R.}~\bibnamefont
  {Aaij}} \emph {et~al.} (\bibinfo {collaboration} {LHCb}),\ }\href {\doibase
  10.1038/s41467-022-30206-w} {\bibfield  {journal} {\bibinfo  {journal}
  {Nature Commun.}\ }\textbf {\bibinfo {volume} {13}},\ \bibinfo {pages} {3351}
  (\bibinfo {year} {2022}{\natexlab{b}})},\ \Eprint
  {http://arxiv.org/abs/2109.01056} {arXiv:2109.01056 [hep-ex]} \BibitemShut
  {NoStop}%
\bibitem [{\citenamefont {Aaij}\ \emph {et~al.}(2014)\citenamefont {Aaij} \emph
  {et~al.}}]{LHCb:2014mir}%
  \BibitemOpen
  \bibfield  {author} {\bibinfo {author} {\bibfnamefont {R.}~\bibnamefont
  {Aaij}} \emph {et~al.} (\bibinfo {collaboration} {LHCb}),\ }\href {\doibase
  10.1103/PhysRevD.90.112004} {\bibfield  {journal} {\bibinfo  {journal} {Phys.
  Rev. D}\ }\textbf {\bibinfo {volume} {90}},\ \bibinfo {pages} {112004}
  (\bibinfo {year} {2014})},\ \Eprint {http://arxiv.org/abs/1408.5373}
  {arXiv:1408.5373 [hep-ex]} \BibitemShut {NoStop}%
\bibitem [{\citenamefont {Cheng}\ \emph {et~al.}(2016)\citenamefont {Cheng},
  \citenamefont {Chua},\ and\ \citenamefont {Zhang}}]{Cheng:2016shb}%
  \BibitemOpen
  \bibfield  {author} {\bibinfo {author} {\bibfnamefont {H.-Y.}\ \bibnamefont
  {Cheng}}, \bibinfo {author} {\bibfnamefont {C.-K.}\ \bibnamefont {Chua}}, \
  and\ \bibinfo {author} {\bibfnamefont {Z.-Q.}\ \bibnamefont {Zhang}},\ }\href
  {\doibase 10.1103/PhysRevD.94.094015} {\bibfield  {journal} {\bibinfo
  {journal} {Phys. Rev. D}\ }\textbf {\bibinfo {volume} {94}},\ \bibinfo
  {pages} {094015} (\bibinfo {year} {2016})},\ \Eprint
  {http://arxiv.org/abs/1607.08313} {arXiv:1607.08313 [hep-ph]} \BibitemShut
  {NoStop}%
\bibitem [{\citenamefont {Chang}\ \emph {et~al.}(2017)\citenamefont {Chang},
  \citenamefont {Chen},\ and\ \citenamefont {Hou}}]{Chang:2017wpl}%
  \BibitemOpen
  \bibfield  {author} {\bibinfo {author} {\bibfnamefont {P.}~\bibnamefont
  {Chang}}, \bibinfo {author} {\bibfnamefont {K.-F.}\ \bibnamefont {Chen}}, \
  and\ \bibinfo {author} {\bibfnamefont {W.-S.}\ \bibnamefont {Hou}},\ }\href
  {\doibase 10.1016/j.ppnp.2017.07.001} {\bibfield  {journal} {\bibinfo
  {journal} {Prog. Part. Nucl. Phys.}\ }\textbf {\bibinfo {volume} {97}},\
  \bibinfo {pages} {261} (\bibinfo {year} {2017})},\ \Eprint
  {http://arxiv.org/abs/1708.03793} {arXiv:1708.03793 [hep-ph]} \BibitemShut
  {NoStop}%
\bibitem [{LHC(2023)}]{LHCb:2023qne}%
  \BibitemOpen
  \href@noop {} {\  (\bibinfo {year} {2023})},\ \Eprint
  {http://arxiv.org/abs/2303.04062} {arXiv:2303.04062 [hep-ex]} \BibitemShut
  {NoStop}%
\bibitem [{\citenamefont {Brayshaw}(1968{\natexlab{a}})}]{PhysRev.176.1855}%
  \BibitemOpen
  \bibfield  {author} {\bibinfo {author} {\bibfnamefont {D.~D.}\ \bibnamefont
  {Brayshaw}},\ }\href {\doibase 10.1103/PhysRev.176.1855} {\bibfield
  {journal} {\bibinfo  {journal} {Phys. Rev.}\ }\textbf {\bibinfo {volume}
  {176}},\ \bibinfo {pages} {1855} (\bibinfo {year}
  {1968}{\natexlab{a}})}\BibitemShut {NoStop}%
\bibitem [{\citenamefont {Brayshaw}(1968{\natexlab{b}})}]{PhysRev.167.1505}%
  \BibitemOpen
  \bibfield  {author} {\bibinfo {author} {\bibfnamefont {D.~D.}\ \bibnamefont
  {Brayshaw}},\ }\href {\doibase 10.1103/PhysRev.167.1505} {\bibfield
  {journal} {\bibinfo  {journal} {Phys. Rev.}\ }\textbf {\bibinfo {volume}
  {167}},\ \bibinfo {pages} {1505} (\bibinfo {year}
  {1968}{\natexlab{b}})}\BibitemShut {NoStop}%
\bibitem [{\citenamefont {Gl\"ockle}(1978)}]{PhysRevC.18.564}%
  \BibitemOpen
  \bibfield  {author} {\bibinfo {author} {\bibfnamefont {W.}~\bibnamefont
  {Gl\"ockle}},\ }\href {\doibase 10.1103/PhysRevC.18.564} {\bibfield
  {journal} {\bibinfo  {journal} {Phys. Rev. C}\ }\textbf {\bibinfo {volume}
  {18}},\ \bibinfo {pages} {564} (\bibinfo {year} {1978})}\BibitemShut
  {NoStop}%
\bibitem [{\citenamefont {Matsuyama}\ and\ \citenamefont
  {Yazaki}(1991)}]{Matsuyama:1991bm}%
  \BibitemOpen
  \bibfield  {author} {\bibinfo {author} {\bibfnamefont {A.}~\bibnamefont
  {Matsuyama}}\ and\ \bibinfo {author} {\bibfnamefont {K.}~\bibnamefont
  {Yazaki}},\ }\href {\doibase 10.1016/0375-9474(91)90464-H} {\bibfield
  {journal} {\bibinfo  {journal} {Nucl. Phys. A}\ }\textbf {\bibinfo {volume}
  {534}},\ \bibinfo {pages} {620} (\bibinfo {year} {1991})}\BibitemShut
  {NoStop}%
\bibitem [{\citenamefont {Konishi}\ \emph {et~al.}(2017)\citenamefont
  {Konishi}, \citenamefont {Morimatsu},\ and\ \citenamefont
  {Yasui}}]{Konishi:2017mpx}%
  \BibitemOpen
  \bibfield  {author} {\bibinfo {author} {\bibfnamefont {A.}~\bibnamefont
  {Konishi}}, \bibinfo {author} {\bibfnamefont {O.}~\bibnamefont {Morimatsu}},
  \ and\ \bibinfo {author} {\bibfnamefont {S.}~\bibnamefont {Yasui}},\
  }\href@noop {} {\  (\bibinfo {year} {2017})},\ \Eprint
  {http://arxiv.org/abs/1705.02569} {arXiv:1705.02569 [hep-ph]} \BibitemShut
  {NoStop}%
\bibitem [{\citenamefont {Bedaque}\ \emph
  {et~al.}(1999{\natexlab{b}})\citenamefont {Bedaque}, \citenamefont {Hammer},\
  and\ \citenamefont {van Kolck}}]{Bedaque:1998km}%
  \BibitemOpen
  \bibfield  {author} {\bibinfo {author} {\bibfnamefont {P.~F.}\ \bibnamefont
  {Bedaque}}, \bibinfo {author} {\bibfnamefont {H.~W.}\ \bibnamefont {Hammer}},
  \ and\ \bibinfo {author} {\bibfnamefont {U.}~\bibnamefont {van Kolck}},\
  }\href {\doibase 10.1016/S0375-9474(98)00650-2} {\bibfield  {journal}
  {\bibinfo  {journal} {Nucl. Phys. A}\ }\textbf {\bibinfo {volume} {646}},\
  \bibinfo {pages} {444} (\bibinfo {year} {1999}{\natexlab{b}})},\ \Eprint
  {http://arxiv.org/abs/nucl-th/9811046} {arXiv:nucl-th/9811046} \BibitemShut
  {NoStop}%
\bibitem [{\citenamefont {Dietz}\ \emph {et~al.}(2022)\citenamefont {Dietz},
  \citenamefont {Hammer}, \citenamefont {K\"onig},\ and\ \citenamefont
  {Schwenk}}]{Dietz:2021haj}%
  \BibitemOpen
  \bibfield  {author} {\bibinfo {author} {\bibfnamefont {S.}~\bibnamefont
  {Dietz}}, \bibinfo {author} {\bibfnamefont {H.-W.}\ \bibnamefont {Hammer}},
  \bibinfo {author} {\bibfnamefont {S.}~\bibnamefont {K\"onig}}, \ and\
  \bibinfo {author} {\bibfnamefont {A.}~\bibnamefont {Schwenk}},\ }\href
  {\doibase 10.1103/PhysRevC.105.064002} {\bibfield  {journal} {\bibinfo
  {journal} {Phys. Rev. C}\ }\textbf {\bibinfo {volume} {105}},\ \bibinfo
  {pages} {064002} (\bibinfo {year} {2022})},\ \Eprint
  {http://arxiv.org/abs/2109.11356} {arXiv:2109.11356 [nucl-th]} \BibitemShut
  {NoStop}%
\bibitem [{\citenamefont {Lindesay}\ and\ \citenamefont
  {Noyes}(1980)}]{Lindesay:1980ib}%
  \BibitemOpen
  \bibfield  {author} {\bibinfo {author} {\bibfnamefont {J.~V.}\ \bibnamefont
  {Lindesay}}\ and\ \bibinfo {author} {\bibfnamefont {H.~P.}\ \bibnamefont
  {Noyes}},\ }\href@noop {} {\bibfield  {journal} {\bibinfo  {journal}
  {SLAC-PUB-2515}\ } (\bibinfo {year} {1980})}\BibitemShut {NoStop}%
\bibitem [{\citenamefont {Lindesay}\ and\ \citenamefont
  {Noyes}(1986)}]{Lindesay:1986kg}%
  \BibitemOpen
  \bibfield  {author} {\bibinfo {author} {\bibfnamefont {J.~V.}\ \bibnamefont
  {Lindesay}}\ and\ \bibinfo {author} {\bibfnamefont {H.~P.}\ \bibnamefont
  {Noyes}},\ }\href@noop {} {\bibfield  {journal} {\bibinfo  {journal}
  {SLAC-PUB-2932}\ } (\bibinfo {year} {1986})}\BibitemShut {NoStop}%
\bibitem [{\citenamefont {Frederico}(1992)}]{Frederico:1992np}%
  \BibitemOpen
  \bibfield  {author} {\bibinfo {author} {\bibfnamefont {T.}~\bibnamefont
  {Frederico}},\ }\href {\doibase 10.1016/0370-2693(92)90661-M} {\bibfield
  {journal} {\bibinfo  {journal} {Phys. Lett. B}\ }\textbf {\bibinfo {volume}
  {282}},\ \bibinfo {pages} {409} (\bibinfo {year} {1992})}\BibitemShut
  {NoStop}%
\bibitem [{\citenamefont {Carbonell}\ and\ \citenamefont
  {Karmanov}(2003)}]{PhysRevC.67.037001}%
  \BibitemOpen
  \bibfield  {author} {\bibinfo {author} {\bibfnamefont {J.}~\bibnamefont
  {Carbonell}}\ and\ \bibinfo {author} {\bibfnamefont {V.~A.}\ \bibnamefont
  {Karmanov}},\ }\href {\doibase 10.1103/PhysRevC.67.037001} {\bibfield
  {journal} {\bibinfo  {journal} {Phys. Rev. C}\ }\textbf {\bibinfo {volume}
  {67}},\ \bibinfo {pages} {037001} (\bibinfo {year} {2003})}\BibitemShut
  {NoStop}%
\bibitem [{\citenamefont {Ydrefors}\ \emph {et~al.}(2017)\citenamefont
  {Ydrefors}, \citenamefont {Alvarenga~Nogueira}, \citenamefont {Gigante},
  \citenamefont {Frederico},\ and\ \citenamefont
  {Karmanov}}]{Ydrefors:2017nnc}%
  \BibitemOpen
  \bibfield  {author} {\bibinfo {author} {\bibfnamefont {E.}~\bibnamefont
  {Ydrefors}}, \bibinfo {author} {\bibfnamefont {J.~H.}\ \bibnamefont
  {Alvarenga~Nogueira}}, \bibinfo {author} {\bibfnamefont {V.}~\bibnamefont
  {Gigante}}, \bibinfo {author} {\bibfnamefont {T.}~\bibnamefont {Frederico}},
  \ and\ \bibinfo {author} {\bibfnamefont {V.~A.}\ \bibnamefont {Karmanov}},\
  }\href {\doibase 10.1016/j.physletb.2017.04.035} {\bibfield  {journal}
  {\bibinfo  {journal} {Phys. Lett. B}\ }\textbf {\bibinfo {volume} {770}},\
  \bibinfo {pages} {131} (\bibinfo {year} {2017})},\ \Eprint
  {http://arxiv.org/abs/1703.07981} {arXiv:1703.07981 [nucl-th]} \BibitemShut
  {NoStop}%
\bibitem [{\citenamefont {Ydrefors}\ \emph {et~al.}(2019)\citenamefont
  {Ydrefors}, \citenamefont {Alvarenga~Nogueira}, \citenamefont {Karmanov},\
  and\ \citenamefont {Frederico}}]{Ydrefors:2019jvu}%
  \BibitemOpen
  \bibfield  {author} {\bibinfo {author} {\bibfnamefont {E.}~\bibnamefont
  {Ydrefors}}, \bibinfo {author} {\bibfnamefont {J.~H.}\ \bibnamefont
  {Alvarenga~Nogueira}}, \bibinfo {author} {\bibfnamefont {V.~A.}\ \bibnamefont
  {Karmanov}}, \ and\ \bibinfo {author} {\bibfnamefont {T.}~\bibnamefont
  {Frederico}},\ }\href {\doibase 10.1016/j.physletb.2019.02.046} {\bibfield
  {journal} {\bibinfo  {journal} {Phys. Lett. B}\ }\textbf {\bibinfo {volume}
  {791}},\ \bibinfo {pages} {276} (\bibinfo {year} {2019})},\ \Eprint
  {http://arxiv.org/abs/1903.01741} {arXiv:1903.01741 [hep-ph]} \BibitemShut
  {NoStop}%
\bibitem [{\citenamefont {Mohseni}\ \emph {et~al.}(2021)\citenamefont
  {Mohseni}, \citenamefont {Chaves}, \citenamefont {da~Costa}, \citenamefont
  {Frederico},\ and\ \citenamefont {Hadizadeh}}]{Mohseni:2021avt}%
  \BibitemOpen
  \bibfield  {author} {\bibinfo {author} {\bibfnamefont {K.}~\bibnamefont
  {Mohseni}}, \bibinfo {author} {\bibfnamefont {A.~J.}\ \bibnamefont {Chaves}},
  \bibinfo {author} {\bibfnamefont {D.~R.}\ \bibnamefont {da~Costa}}, \bibinfo
  {author} {\bibfnamefont {T.}~\bibnamefont {Frederico}}, \ and\ \bibinfo
  {author} {\bibfnamefont {M.~R.}\ \bibnamefont {Hadizadeh}},\ }\href {\doibase
  10.1016/j.physletb.2021.136773} {\bibfield  {journal} {\bibinfo  {journal}
  {Phys. Lett. B}\ }\textbf {\bibinfo {volume} {823}},\ \bibinfo {pages}
  {136773} (\bibinfo {year} {2021})},\ \Eprint
  {http://arxiv.org/abs/2111.02015} {arXiv:2111.02015 [nucl-th]} \BibitemShut
  {NoStop}%
\bibitem [{\citenamefont {Frederico}\ and\ \citenamefont
  {Ydrefors}(2021)}]{Frederico:2021wej}%
  \BibitemOpen
  \bibfield  {author} {\bibinfo {author} {\bibfnamefont {T.}~\bibnamefont
  {Frederico}}\ and\ \bibinfo {author} {\bibfnamefont {E.}~\bibnamefont
  {Ydrefors}},\ }\href {\doibase 10.1007/s00601-021-01594-4} {\bibfield
  {journal} {\bibinfo  {journal} {Few Body Syst.}\ }\textbf {\bibinfo {volume}
  {62}},\ \bibinfo {pages} {8} (\bibinfo {year} {2021})}\BibitemShut {NoStop}%
\bibitem [{\citenamefont {Hansen}\ \emph {et~al.}(2021)\citenamefont {Hansen},
  \citenamefont {Brice\~no}, \citenamefont {Edwards}, \citenamefont {Thomas},\
  and\ \citenamefont {Wilson}}]{Hansen:2020otl}%
  \BibitemOpen
  \bibfield  {author} {\bibinfo {author} {\bibfnamefont {M.~T.}\ \bibnamefont
  {Hansen}}, \bibinfo {author} {\bibfnamefont {R.~A.}\ \bibnamefont
  {Brice\~no}}, \bibinfo {author} {\bibfnamefont {R.~G.}\ \bibnamefont
  {Edwards}}, \bibinfo {author} {\bibfnamefont {C.~E.}\ \bibnamefont {Thomas}},
  \ and\ \bibinfo {author} {\bibfnamefont {D.~J.}\ \bibnamefont {Wilson}}
  (\bibinfo {collaboration} {Hadron Spectrum}),\ }\href {\doibase
  10.1103/PhysRevLett.126.012001} {\bibfield  {journal} {\bibinfo  {journal}
  {Phys. Rev. Lett.}\ }\textbf {\bibinfo {volume} {126}},\ \bibinfo {pages}
  {012001} (\bibinfo {year} {2021})},\ \Eprint
  {http://arxiv.org/abs/2009.04931} {arXiv:2009.04931 [hep-lat]} \BibitemShut
  {NoStop}%
\bibitem [{\citenamefont {Hansen}\ and\ \citenamefont
  {Sharpe}(2016)}]{Hansen:2016fzj}%
  \BibitemOpen
  \bibfield  {author} {\bibinfo {author} {\bibfnamefont {M.~T.}\ \bibnamefont
  {Hansen}}\ and\ \bibinfo {author} {\bibfnamefont {S.~R.}\ \bibnamefont
  {Sharpe}},\ }\href {\doibase 10.1103/PhysRevD.93.096006} {\bibfield
  {journal} {\bibinfo  {journal} {Phys. Rev. D}\ }\textbf {\bibinfo {volume}
  {93}},\ \bibinfo {pages} {096006} (\bibinfo {year} {2016})},\ \bibinfo {note}
  {[Erratum: Phys.Rev.D 96, 039901 (2017)]},\ \Eprint
  {http://arxiv.org/abs/1602.00324} {arXiv:1602.00324 [hep-lat]} \BibitemShut
  {NoStop}%
\bibitem [{\citenamefont {Blanton}\ \emph {et~al.}(2019)\citenamefont
  {Blanton}, \citenamefont {Romero-L\'opez},\ and\ \citenamefont
  {Sharpe}}]{Blanton:2019igq}%
  \BibitemOpen
  \bibfield  {author} {\bibinfo {author} {\bibfnamefont {T.~D.}\ \bibnamefont
  {Blanton}}, \bibinfo {author} {\bibfnamefont {F.}~\bibnamefont
  {Romero-L\'opez}}, \ and\ \bibinfo {author} {\bibfnamefont {S.~R.}\
  \bibnamefont {Sharpe}},\ }\href {\doibase 10.1007/JHEP03(2019)106} {\bibfield
   {journal} {\bibinfo  {journal} {JHEP}\ }\textbf {\bibinfo {volume} {03}},\
  \bibinfo {pages} {106} (\bibinfo {year} {2019})},\ \Eprint
  {http://arxiv.org/abs/1901.07095} {arXiv:1901.07095 [hep-lat]} \BibitemShut
  {NoStop}%
\bibitem [{\citenamefont {Brice\~no}\ \emph {et~al.}(2018)\citenamefont
  {Brice\~no}, \citenamefont {Hansen},\ and\ \citenamefont
  {Sharpe}}]{Briceno:2018mlh}%
  \BibitemOpen
  \bibfield  {author} {\bibinfo {author} {\bibfnamefont {R.~A.}\ \bibnamefont
  {Brice\~no}}, \bibinfo {author} {\bibfnamefont {M.~T.}\ \bibnamefont
  {Hansen}}, \ and\ \bibinfo {author} {\bibfnamefont {S.~R.}\ \bibnamefont
  {Sharpe}},\ }\href {\doibase 10.1103/PhysRevD.98.014506} {\bibfield
  {journal} {\bibinfo  {journal} {Phys. Rev. D}\ }\textbf {\bibinfo {volume}
  {98}},\ \bibinfo {pages} {014506} (\bibinfo {year} {2018})},\ \Eprint
  {http://arxiv.org/abs/1803.04169} {arXiv:1803.04169 [hep-lat]} \BibitemShut
  {NoStop}%
\end{thebibliography}%

\clearpage
\onecolumngrid


\section*{Supplemental Material}

\subsection{Three-body integral equations}
\label{app:suppiso}
Here, we review the key expressions needed to solve for the three-body scattering amplitude. We employ the relativistic integral equations defined in Ref.~\cite{Hansen:2015zga} for the $\Mc_{3}$. We note that similar three-body models were studied using non-relativistic Faddeev equations \citep{PhysRev.176.1855, PhysRev.167.1505, PhysRevC.18.564, Matsuyama:1991bm, Hyodo:2013zxa, Konishi:2017mpx} and non-relativistic EFTs \citep{Bedaque:1998kg, Bedaque:1998km, Dietz:2021haj}. First attempts to understand the system of three bosons in the relativistic framework have been made in Refs.~\citep{Lindesay:1980ib, Lindesay:1986kg}.  However, their result disagreed with the subsequent Light-Front calculation \citep{Frederico:1992np} obtained in an equivalent regularization scheme. Its redefinition \citep{PhysRevC.67.037001} led to an approximate agreement between two approaches but also implied the unphysical behavior of the trimer mass becoming imaginary at finite values of $a$. We do not observe any indication of the Thomas collapse reported in this work. One can find additional studies in Refs.~\citep{Ydrefors:2017nnc, Ydrefors:2019jvu, Mohseni:2021avt, Frederico:2021wej} and a summary of the situation in Ref.~\citep{Naidon:2016dpf}.

Strictly speaking, in this article, we work with the pair-spectator amplitude $\Mc^{(u,u)}_3$, where specific spectators are chosen. The superscript emphasizes that the initial and final states are ``unsymmetrized". It becomes the genuine $3\varphi \to 3 \varphi$ amplitude $\Mc_3$ only after symmetrization with respect to different spectator choices~\citep{Hansen:2015zga}. The amplitude is expressed as a sum of two terms,
    \begin{align}
    \label{eq:Def.M3uu}
    \Mc^{(u,u)}_3(p,k) = \Dc^{(u,u)}(p,k) + \Mc_{\text{df},3}^{(u,u)}(p,k) \, .
    \end{align}
The first term is called the \emph{ladder} amplitude and describes the scattering driven by one-particle exchanges, with the three-body couplings ``turned off''. The second term depends on the three-body $K$ matrix, $\Kc_{3}$, introduced in the main text. Compared to Refs.~\citep{Hansen:2014eka, Hansen:2015zga} we drop the ``df'' label from this object to simplify the notation. We also drop the ${(u,u)}$ superscript from the $\Mc_3$, $\Dc$, and $\Mc_{\text{df},3}$ remembering these are ``unsymmetrized" objects.

Following Refs.~\cite{Jackura:2020bsk, Hansen:2020otl, Dawid:2023jrj}, we perform the partial-wave projection, setting all relevant angular momenta to zero. The equation for the partial-wave projected version of $\Dc$ is,
    \begin{align}
    \label{eq:Ddef_v2}
    \Dc(p,k) 
    = 
    - \Mc_2( p) \, G(p,k) \, \Mc_2( k) 
    - \Mc_2( p) \int\limits_0^{k_{\rm max}} \! \frac{{\rm d} k' \, k'^2}{(2 \pi)^2 \omega_{k'}}  \, G(p,k') \,  \Dc(k',k)  \, .
    \end{align}
All objects in the above equation implicitly depend on the total energy variable, $E$. The one-particle exchange (OPE) amplitude $G$ is the boson exchange propagator,
    \begin{align}
    G(\p,\k)
    =
    \frac{ H( p, k ) }
    {(E - \omega_p - \omega_k)^2 - (\p + \k)^2 - m^2 + i \epsilon} \, ,
    \end{align}
which, after the partial-wave projection, becomes,
    \begin{align}
    \label{eq:Gs_proj}
    G(p, k)
    & = - \frac{H(p, k)}{4pk} \, \log\left( \frac{z(p, k) + i \epsilon - 2pk}{z(p, k) + i \epsilon + 2pk} \right) \, .
    \end{align}
Here $z(p, k) = (E-\omega_{k} - \omega_p)^2 - k^2 - p^2 - m^2$ and $H(p,k)$ is a cutoff function. In previous work~\cite{Jackura:2020bsk, Dawid:2023jrj}, we employed smooth regularization choice. Here, we exclusively consider hard cutoff functions that are defined to be equal to $1$ in the range of integration and $0$ otherwise. The maximal value the momentum takes is given by $k_{\rm max} = \sqrt{E^2 - m^2}/{2 E}$ which corresponds to minimal $\varepsilon_{k'} = 0$ of the pair in the intermediate three-body state.

The second term of Eq.~\eqref{eq:Def.M3uu}, amplitude $\Mc_{\text{df},3}$, can be written as a function of $\Dc$. Here we only consider the simplest scenario (the so-called isotropic approximation) where the three-body $K$ matrix depends solely on the total energy of the system, $\Kc_3(p,k) \equiv \Kc_3(E)$. The three-body matrix $\Kc_3$ is practically an unconstrained function of $E$ that can consist of a polynomial of arbitrary order and a sum of poles. In this case, $\Mc_{\text{df},3}$ takes a form,
    \begin{align}
    \label{eq:M3df}
    \Mc_{\text{df},3}(p,k)
    &= \Lc(p) \, \frac{1}{ \Kc_{3}^{-1} + F_3^{\infty}} \, \Lc(k) \, ,
    \end{align}
where the ``endcap" functions describing two-body rescattering contributions are,
    \begin{align}
    \label{eq:Lc}
    \Lc(p) &= \frac{1}{3} - \Mc_2(p) \rho(p) 
    - \int_{k'} \Dc(p,k') \rho(k') \, . 
    \end{align}
The three-body kinematic function, $F_3^\infty$, describes the effects of propagation of on-shell three-body states interacting either within two-body subchannels or via one-particle exchanges. It is expressed in terms of $\Mc_2$ and $\Dc$,
    \begin{align}
    \label{eq:F3}
    F_3^\infty = \int_{k'} \rho(k') \, \Lc(k')
    = \int_{k'} \rho(k') 
    \left(\frac{1}{3} - \Mc_2(k') \rho(k') \right)
    - \int_{p'}\int_{k'}
    \rho(p') \Dc(p',k') \rho(k') \, . 
    \end{align}
Finally, the phase space function, $\rho$, is defined as,
    \begin{equation}
    \rho(k) =  \, 
    \frac{-i}{16 \pi \varepsilon_k} \sqrt{ \varepsilon_k^2/4 - m^2 }  \, .
    \end{equation}  
Note that this differs slightly from what was presented in, for example, Refs.~\citep{Hansen:2014eka, Hansen:2015zga}, where the authors used the smooth cutoff function. As a result, the phase space was proportional to the cutoff function. We also dropped the ``3" subscript from the definition of the phase space to simplify the notation.


\subsection{Cancellation of $\Kc_3=0$ poles}

We note that separation into long- and short-range forces is ambiguous since $\Kc_3$ is a regularization-scheme dependent object~\citep{Jackura:2022gib}. Cutoff dependence of $\Dc$ is compensated by the changes of $\Mc_{\text{df},3}$ under variations of $\Kc_3$, assuring the scattering amplitude is independent of the regularization choice. In the main text we consider the case $\Kc_3 = 0$, for which $\Mc_{\text{df},3}=0$ and $\Mc_3 = \Dc$. The purpose of this and the subsequent section is to argue that this particular choice does not affect our main conclusions. In particular, we would like to verify that near the unitarity, $\Mc_3$ develops Efimov bound states for a non-zero three-body $K$ matrix.

From Eq.~\eqref{eq:M3df}, we see that $\Mc_{\text{df},3}$ contributes poles to the full three-body amplitude whenever,
    \begin{align}
    \label{eq:pole-condition}
    \Kc_{3}^{-1} + F_3^{\infty} \Big|_{E=E_n} = 0 \, ,
    \end{align}
As a first step, we would like to show that all the poles of $\Mc_3$ are those described by the above equation. Although it is evident that Eq.~\eqref{eq:pole-condition} does indeed describe some states, it is not immediately clear that it governs all of them. In particular, the first term in Eq.~\eqref{eq:Def.M3uu}, $\Dc$, can contribute additional poles to the full three-body amplitude.

Assume $\Dc$ has a pole at position $E_{n,0}$, i.e., it can be expanded as,
    \begin{align}
    \Dc(p, k)
    &= -\frac{\Gamma_{n,0}(p) \Gamma_{n,0}(k)}{E^2-E_{n,0}^2} + \Oc\!\left(E^0\right) \, ,
    \end{align}
where the subscript ``$0$" is meant to emphasize that this corresponds to the $\Kc_3=0$ limit of $\Mc_{3}$. We note that when $\Dc$ has a pole in $E^2$, from Eqs.~\eqref{eq:Lc},~\eqref{eq:F3} it is clear that so do $\Lc$ and $F_3^{\infty}$. In its vicinity, these functions take the form,
    \begin{align}
    \label{eq:Lc_pole}
    \Lc(p) &= 
    \frac{\Gamma_{n,0}(p) }{E^2-E_{n,0}^2} \int_{k'} \Gamma_{n,0}(k') \rho(k') \, 
    + \Oc\!\left(E^0\right),
    \\
    \label{eq:F3_pole}
    F_3^{\infty}&= 
    \frac{1 }{E^2-E_{n,0}^2} \left[\int_{k'} \Gamma_{n,0}(k') \rho(k') \right]^2 \, 
    + \Oc\!\left(E^0\right) \, .
    \end{align}
Using these expansions in Eq.~\eqref{eq:M3df} we find,
    \begin{align}
    \Mc_{{\rm df},3}(p, k)
    &= \frac{\Gamma_{n,0}(p) \Gamma_{n,0}(k)}{E^2-E_{n,0}^2} + \Oc\!\left(E^0\right) ,
    \end{align}
which exactly cancels the pole contributions from $\Dc$ in Eq.~\eqref{eq:Def.M3uu}. As a result, the only poles present in $\Mc_3$ are those given by Eq.~\eqref{eq:pole-condition}.


\subsection{Behavior of $F_3^\infty$}
\label{sec:f_infty}

\begin{figure}[t]
    \centering
    \includegraphics[width=0.8\textwidth, trim = {0 0 0 0}, clip]{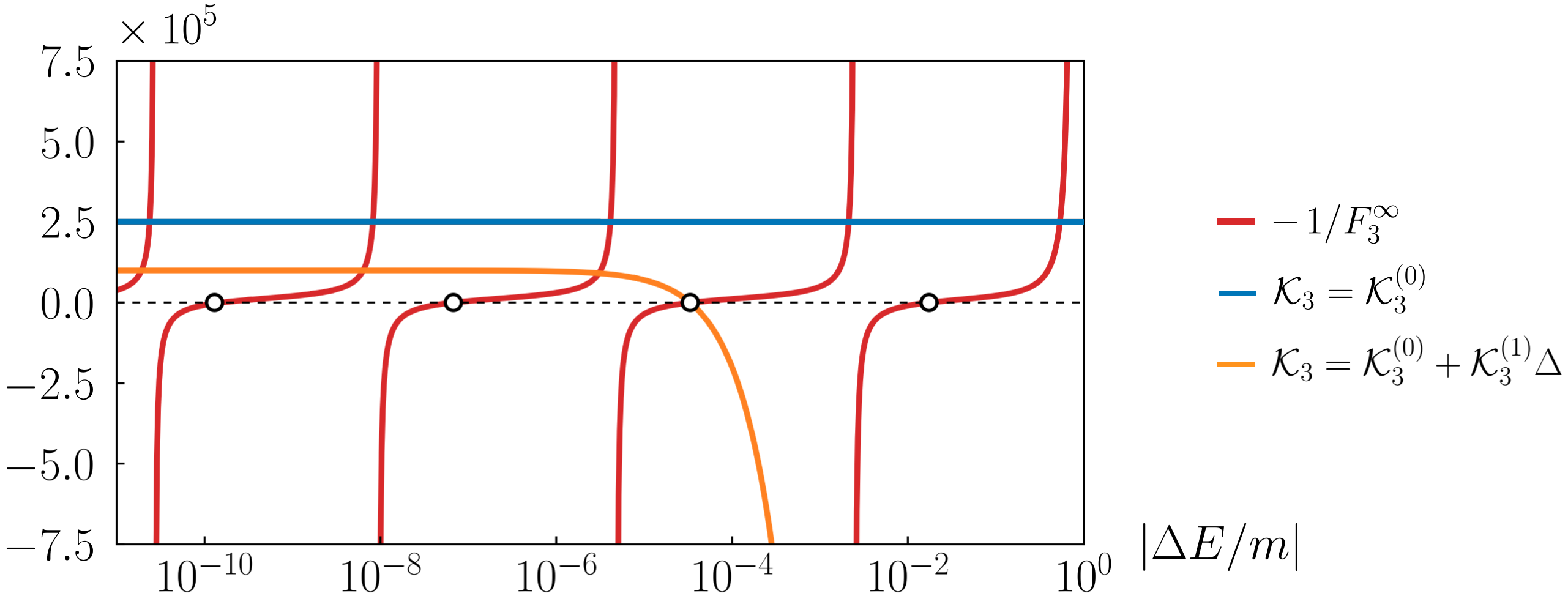}
    \caption{Inverse $-1/F_3^\infty$ as a function of $\Delta E$, for $ma=10^6$. Pole positions of $\Mc_3$ are determined by finding values of $\Delta E$ at which $\Kc_3$ crosses the red curves. Two typical forms of $\Kc_3$ are shown: a constant $\Kc_3^{(0)}/m = 2.5 \cdot 10^5$ in brown, and a NLO threshold-expanded isotropic version in orange~\citep{Hansen:2016fzj, Blanton:2019igq}. For the latter: $\Kc_3^{(0)}/m = 10^5$, $\Kc_3^{(1)}/m = 4.5 \cdot 10^{9} $, and $\Delta = (E^2 - (E_{\text{thr.}}^{(3 \varphi)})^2)/(E_{\text{thr.}}^{(3 \varphi)})^2$. Dots represent binding energies in the limit $\Kc_3 = 0$ for which $\Mc_3 = \Dc$.}
    \label{fig:k3-scaling}
\end{figure}

Equation~\eqref{eq:pole-condition} describes states appearing in the $\Kc_3=0$ limit since the $F_3^{\infty}$ function has the same poles as $\Dc$, i.e., it diverges whenever $E=E_n$. Close to unitarity, by solving Eq.~\eqref{eq:3body_inteq} numerically, we find that $F_3^{\infty}$ becomes a log-periodic function of binding energy, $\Delta E$. Moreover, the sign of the $F_{3}^\infty$'s first derivative is constrained below the threshold. These features are shown in Fig.~\ref{fig:k3-scaling}. Thus, in the $\Kc_3=0$ case, the bounds states obey the discrete scaling symmetry characteristic of Efimov's phenomenon. As we go to higher $a$, poles of $F_3^\infty$ accumulate near the energy threshold, and the energy gap between them becomes increasingly small.

In that region, regardless of its functional form, $\Kc_3$ is to a good approximation constant\footnote{Except for instances when it has a pole exactly at the threshold.}, see an orange line in Fig.~\ref{fig:k3-scaling} for an example. Therefore, near the threshold, we recover the Efimov scaling for any model of $\Kc_3$—an effect ensuring the universality of our result. It confirms the cutoff independence of this behavior at unitarity since various models of the three-body $K$ matrix correspond to different regularization choices.

\subsection{Analytic continuation}
\label{app:anal_cont}

Here, we briefly describe the analytic continuation of the integral equation defining the $\Dc(p,k)$ amplitude.
We wish to extend it from the real axis to the complex energy, which is achieved by the method of the integration contour deformation. Generalization of the integration interval from a straight real line to a complex curved path is motivated by the presence of movable (energy-dependent) singularities of $G(p,k')$, $\Mc_2(k')$, and $\Dc(k',k)$ that can cross real $k'$ momentum axis when $E^2$ becomes complex-valued. These singularities must be avoided at each value of $E^2$ and $p$; otherwise, the final result becomes contour-dependent, invalidating the uniqueness of the analytic continuation. 

The main focus of this article, i.e., identification of the trimer states as poles of the amplitude, requires a solution of the integral equation for real energies below the ${\varphi b}$ and ${3 \varphi}$ thresholds and for complex energy in the unphysical Riemann sheets associated with these two branch cuts. We first briefly summarize methods employed in Refs.~\citep{Jackura:2020bsk, Dawid:2023jrj} to analyze the $\varphi b$ system and only then describe the continuation of the amplitude through the three-body cut. We focus on the $\Kc_3 = 0$ case for clarity of the presentation. Once we compute the ladder amplitude from Eq.~\eqref{eq:Ddef_v2}, we can obtain the full $\Mc_{3}$ amplitude by utilizing formulas provided in Sec.~\ref{app:suppiso} of this supplement.

\subsubsection{The dimer-particle threshold}

In Ref.~\citep{Jackura:2022gib}, we explained how the scattering amplitude between a two-body bound state and a spectator, $\Mc_{\varphi b}$, is obtained from the three-body scattering amplitude using the LSZ reduction formula. The amplitude $\Dc(p,k)$ has poles at values of the pairs' energies squared, $\varepsilon_p^2, \varepsilon_k^2$, equal to that of the bound state, $m_b^2$. The residue of the three-body amplitude at these poles is proportional to $\Mc_{\varphi b}$. Explicitly, expanding $\Dc(p,k)$ in their vicinity, we find,
    \begin{align}
    \label{eq:LSZ}
    \Dc(p,k) =
    \frac{g^2 \, \Mc_{\varphi b}(E)}{(\varepsilon_p^2-m_b^2)(\varepsilon_k^2-m_b^2)} + \cdots \, ,
    \end{align}
where $g^2 = 128\pi m_b/a$. Continuing the external spectator momenta, $p$ and $k$, to the value corresponding to the two-body bound state pole, $q_b$, 
    \beq
    q_b = \frac{\lambda^{1/2}(E^2, m_b^2, m^2)}{2 E} \, ,
    \eeq
brings $\varepsilon_p^2, \varepsilon_k^2 \to m_b^2$, allowing to extract the residue from the above formula. Here $\lambda(x,y,z) = x^2 + y^2 + z^2 - 2xy - 2xz - 2yz$ is the K\"all\'en triangle function.

\begin{figure}[t]
    \centering
    \includegraphics[width=0.99\textwidth, trim = {0 0 0 0}, clip]{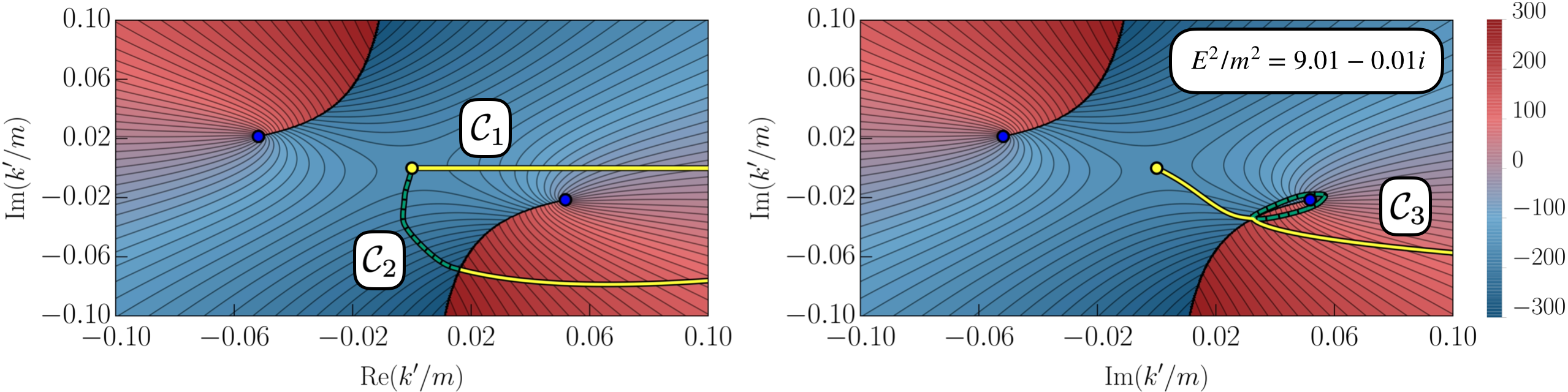}
    \caption{Imaginary part of $\Mc_2(k')$ in the complex $k'$ plane for $ma=-6$ and $s/m^2 = 9.01 - 0.01i$. Black lines represent branch cuts, and blue dots are branch points. Three example integration contours are shown. On the left panel, $\Cc_1$ (yellow, straight line) is the original integration interval of Eq.~\eqref{eq:3body_inteq} defining the solution, $\Mc_3(p,k)$, on the physical Riemann sheet of the complex $E^2$ plane. To continue the three-body amplitude through the three-body cut, one deforms $\Cc_1 \to \Cc_2$. Amplitude $\Mc_2$ in the integration kernel must be evaluated on its second sheet when $k'$ belongs to the green, dashed piece of $\Cc_2$. On the right panel, we show an example contour, $\Cc_3$, defining the ladder amplitude on the +2 unphysical Riemann sheet of the $E^2$ plane.}
    \label{fig:5}
\end{figure}

The trimer bound and virtual states are poles of the $\Mc_{\varphi b}$ amplitude below the $\varphi b$ threshold that lie on the first and second Riemann sheets, respectively. From the two-body $S$ matrix unitarity, the analytic continuation of the amplitude to the unphysical Riemann sheet, $\Mc_{\varphi b}^{\text{II}}$, is known explicitly,
    \beq
    \label{eq:phi-b-second-sheet}
    \Mc_{\varphi b}^{\text{II}}(E) = \frac{\Mc_{\varphi b}(E)}{1 + 2 i \rho_{\varphi b}(E) \Mc_{\varphi b}(E) } \, ,
    \eeq
where $\rho_{\varphi b}(E)$ is a $\varphi b$ phase-space factor,
    \beq
    \rho_{\varphi b}(E) = \frac{q_b}{8 \pi E} \, .
    \eeq
Thus, the knowledge of $\Mc_{\varphi b}$ on the first Riemann sheet is sufficient to recover both the positions of three-body bound and virtual states.

The main complication in extending the three-body amplitude below the dimer-particle threshold $E_{\text{thr}}^{(\varphi b)}$ is the logarithmic branch cut of the homogeneous and inhomogeneous terms of the integral equation in the $p$ variable. It originates from the OPE partial-wave projected propagator, Eq.~\eqref{eq:Gs_proj}, present in both of these terms and thus is an ``inherited" singularity of the $\Dc$ amplitude, appearing in its left-hand argument~\cite{Dawid:2023jrj}. At a fixed value of momentum $k$ and total energy, the inhomogeneous term of the equation contributes a fixed cut that has to be avoided by the deformed integration contour. In particular, for external momentum $k=q_b$ and $E < E_{\text{thr}}^{(\varphi b)}$, the cut may take a complicated shape resembling a circle, see Fig.~4 in Ref.~\citep{Dawid:2023jrj}.

The OPE amplitude in the homogeneous part of the equation is evaluated at complex momenta $k'$ along the contour chosen to avoid this fixed quasi-circular shape. Since the right-hand argument of $G(p,k')$ takes infinitely many values, this amplitude contributes a whole set of singularities to the left-argument dependence of the ladder amplitude that we call a \textit{domain of non-analyticity}. Note that this area of the complex $k'$ plane is defined by the specific integration path chosen. The deformed contour must detour this region of the complex plane, a condition we call self-consistency of the integration contour. Only self-consistent integration paths define proper analytic continuation of the integral equation solution. An example domain of non-analyticity is found in Fig.~8 of Ref.~\citep{Dawid:2023jrj}.

We refer the reader interested in more details to this work. Once we verified that all singularities of the integration kernel and of the ladder amplitude $\Dc(k',k)$ are bypassed, we may attempt to solve the integral equation numerically along the deformed contour, as described in App.~C therein.

\subsubsection{Three-particle threshold}

Analytic continuation of the ladder amplitude through the three-body (logarithmic) cut to the associated unphysical Riemann sheet requires further discussion that is new to this work.

\begin{figure}[t]
    \centering
    \includegraphics[width=0.99\textwidth, trim = {0 0 0 0}, clip]{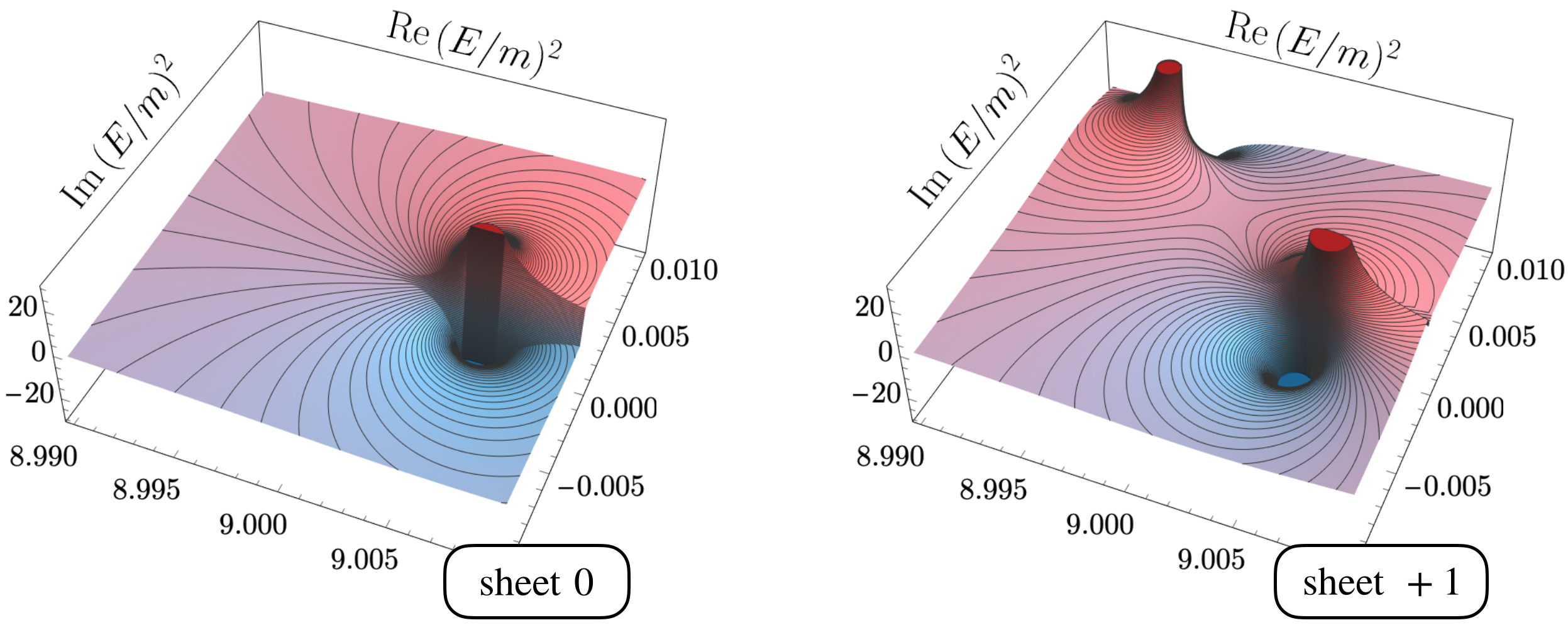}
    \caption{Imaginary part of the amplitude $m^2 \, d(p,k)$ in the complex $E^2$ plane for $ma=-8.1$ and $\varepsilon_p = \varepsilon_k = 2 m^2$. The left panel represents the solution on sheet 0, while the right panel is obtained via the analytic continuation to the nearest unphysical sheet.}
    \label{fig:6}
\end{figure}

To simplify the discussion, let us amputate the external two-body amplitudes from the integral equation, obtaining,
    \beq
    \label{eq:ddef}
    d(p,k) 
    = 
    - G(p,k)
    - \int\limits_0^{k_{\rm max}} \! \frac{{\rm d} k' \, k'^2}{(2 \pi)^2 \omega_{k'}}  \, G(p,k') \, \Mc_2( k') \, d(k',k)  \, ,
    \eeq
where $\Dc(p,k) = \Mc_2(p) \, d(p,k) \, \Mc_2(k)$. In addition to deciding whether the behavior of the singularities of the OPE amplitude demands integration contour deformation, one has to consider the singularities of the two-body amplitude, $\Mc_2$, in the integration kernel. As described in the main text, the singularities of the amplitude in the $E$ variable emerge from the collision of the integration end point with the singularities of $\Mc_2$ in the integration kernel. The three-body right-hand cut of the three-body amplitude arises when the right-hand, two-body branch point of $\Mc_2$ at 
    \beq
    k' = \frac{\lambda^{1/2}(E^2,m^2,(2m)^2)}{2 E}
    \eeq
coincides with the lower limit of integration, $k' = 0$. A particular choice of the integration contour behavior around this branch point determines the Riemann sheet of the obtained integral equation solutions; see Fig.~\ref{fig:5}.

Namely, when the integration contour circumvents the branch point from the top, and the $\Mc_2$ is evaluated exclusively on its first Riemann sheet, the integral equation defines the amplitude on the first (physical) Riemann sheet above the three-particle threshold. It is not the only available option. The integration contour can pass the branch point from the bottom. In this case, the three-body is computed on its first unphysical (+1) Riemann sheet of the complex $E$ plane. Since there is a cut extending from the $\varepsilon_{k'} = (2m)^2$ point, to properly define the analytic continuation of the three-body amplitude, one must add a relevant discontinuity function to $\Mc_2$, when evaluating the integration kernel in this case, as indicated by the dashed green line in Fig.~\ref{fig:5}. 

Finally, when the integration contour encircles the branch point $n$ times, as shown in the right panel of Fig.~\ref{fig:5}, the amplitude is computed on the $(1+n)^{\rm th}$ Riemann sheet, using the notation of Fig.~\ref{fig:sheets}. We remind the reader that the three-body cut is logarithmic, i.e., contrary to algebraic discontinuities such as in the square root function, it is associated with infinitely many Riemann sheets. Winding around the branch point generates them in the integral-equation representation of the three-body amplitude. 

In Fig.~\ref{fig:6} we present example solution of Eq.~\eqref{eq:ddef} for $ma=-8.1$ and external momenta $p=k$ chosen such that $\varepsilon_k^2 = \varepsilon_p^2 = 2 m^2$. The solution on the physical Riemann sheet (left panel) exhibits a clear resonance ``bump" above the real axis, hinting at a pole on the nearest unphysical sheet. The right panel presents the result of the analytic continuation to the +1 branch of the amplitude, which reveals a ground state pole at position $E_1^2/m^2 \approx 9.0050 - 0.0012i$. It is a pole indicated by the red color in Figs.~\ref{fig:kappa_trajectories} and~\ref{fig:3-resonance} in the main text. Another pole, which we discuss in the next section, is found in the upper half plane, at position $E/m^2 \approx 8.9944 + 0.0087i$. We also observe another ``bump" in the amplitude above the real axis on sheet +1, a hint that another pole could be expected on sheet +2.

\subsubsection{Three-particle threshold --- sheets $\geq +2$ }

Unfortunately, we found it practically impossible to define integration contours of the type presented in the right panel of Fig.~\ref{fig:5}, that would be self-consistent, i.e., that would avoid additional singularities generated by the one-particle exchange amplitude. We note that the domain of non-analyticity always seems to contain the branch point of $\Mc_2 (k')$, meaning that it is unattainable to surround encircle it with a loop-like integration contour without crossing it. A rigorous solution to the problem of non-consistency of the integration paths required to extend the three-body amplitude to sheets $\geq +2$ needs additional study and is beyond the scope of this paper. As we note in the main text, continuation to the higher Riemann sheets is most probably the solution to the ``missing pole" problem and thus will be of practical interest in the future study of the three-body problem.

\begin{figure}[t]
    \centering
    \includegraphics[width=0.99\textwidth, trim = {0 0 0 0}, clip]{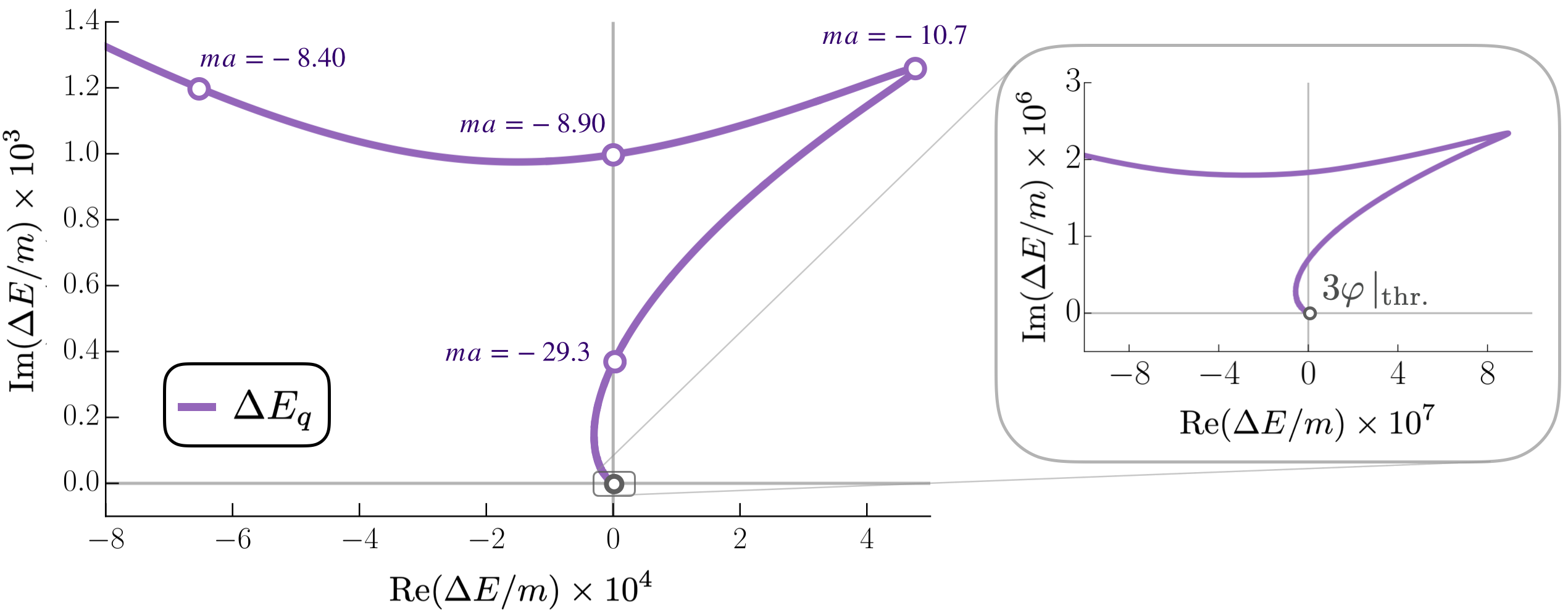}
    \caption{Evolution of the ``quirky" pole in the upper half-plane of the +1 Riemann sheet. The pole approaches the threshold from complex infinity. As $|ma|$ decreases, its trajectory forms a fractal pattern with rescaling quotient converging to $\lambda^2$.}
    \label{fig:7}
\end{figure}


\subsection{Additional pole on sheet +1}

In addition to the regular Efimov resonances in the +1 Riemann sheet of the complex $E^2$ plane, we also find a new, ``quirky" state in the upper half-plane, $\im E^2 > 0$. This trimer pole is visible on the right panel of Fig.~\ref{fig:6}, in the upper left corner of the frame. We trace its trajectory with changing two-body scattering length and present it in Fig.~\ref{fig:7}, denoting its binding energy as $\Delta E_q$. We observe as it approaches the three-body threshold from a deep region of the complex plane, supposedly the complex infinity. Interestingly, as the magnitude of $ma$ increases, it bypasses the three-body branch point and turns back. This pattern keeps repeating for arbitrarily high values of $ma$. The fractal trajectory it forms has a self-similarity factor converging to the Efimov's constant, $\lambda^2$.

We find this state puzzling, and at this point we can not draw any conclusions on it nature. The pole approaches the three-body branch point in the unitarity limit and could play a role in the accumulation of bound states poles at the threshold. It could be a physically interesting phenomenon or simply an artifact of the relativistic integral equations. At this time, neither option can be ruled out, and further investigations are needed. 

\begin{figure}[t]
    \centering
    \includegraphics[width=0.99\textwidth, trim = {0 0 0 0}, clip]{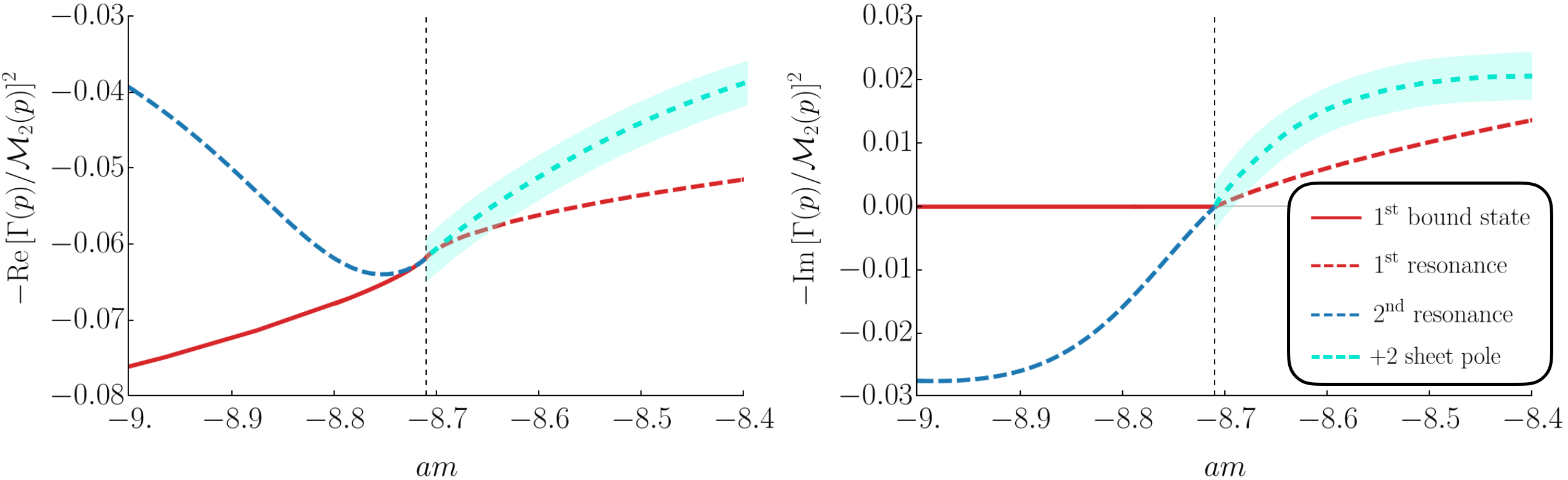}
    \caption{The $ma$ dependence of the normalized trimer residues. Real (left) and imaginary (right) parts of the bound state residue is shown with a red, solid line. Residues of resonance poles are presented as dashed lines. The $+2$ sheet pole, obtained through the extrapolation procedure, is shown with an error band estimate. Around $ma = -8.71$ the poles approach the threshold and all residues converge to the same value, around $-0.061$.}
    \label{fig:8}
\end{figure}


\subsection{Matching of pole residues at the three-body threshold}

As explained in the main text, by studying evolution of the trimer poles, we found that the $n^{\text{th}}$ resonance on the +1 sheet reaches the $3 \varphi$ threshold at the same value of the two-body scattering length at which the $n^{\text{th}}$ bound state emerges on the physical energy plane, and the $(n+1)^{\text{th}}$ resonance appears on the +1 sheet. For instance, for $n=1$ this occurs around $ma=-8.71$, while for $n=2$ around $ma=-203$. Moreover, even though the residues behave in a seemingly unrelated manner away from the $E_{\text{thr}}^{(3\varphi)}$, they converge to the same number at the value of $ma$ where the transition between Riemann sheets occurs.

In Fig.~\ref{fig:8}, we present the numerical evidence of this surprising behavior for the $n=1$ case. We fix $p=k$ by choosing $\varepsilon_p=\varepsilon_k = 2m^2$ and compute the residues of the 1st bound state and the 1st and 2nd resonances. We normalize them by $\Mc_2(p)^2$ to cancel enhancement from the two-body rescatterings in the final and initial pairs.

Although we can not analytically continue the amplitude to the Riemann sheets higher than the first unphysical one, for some values of the scattering length, we see its enhancement above the real axis on the $+1$ sheet. It is reminiscent of the narrow resonant ``bump" and suggests the existence of a pole in the lower half-plane of the $+2$ sheet. Working under the assumption that this pole is located near the threshold at $ma=-8.71$, we perform a Breit-Wigner fit to the $\Dc(p,p)$ amplitude $i \epsilon$ above the real energy axis to extract its position and residue. We use the model,
    \beq
    \Dc(p,p) = - \frac{\Gamma_m^2}{ E^2 + i \epsilon - E_m^2 } + f(E) \, ,
    \eeq
where we again fix $p=k$ such that $\varepsilon_k = \varepsilon_p = 2m^2$. The $\Gamma_m^2$ is a complex residue, $E_m$ is the pole position of the modeled trimer, and $f(E)$ is a background quadratic function. We perform several fits, differing by the value of $\epsilon$ and the form of $f(E)$. Our model depends, in total, on four real parameters when $f(E) = 0$ up to ten for $f(E) = c_0 + c_1 E + c_3 E^2$, where all $c_i$ are complex numbers. At each $ma$, we find little variability in values of $\Gamma_m^2$ and $E_m^2$ between different fits and take the largest difference between any two of them as the error estimate.

The pole position extracted in this way approaches the three-body threshold as $ma \to -8.71$. In Fig.~\ref{fig:8}, we also show evidence that the residue of this pole converges to the same value as the residues of the remaining ones, supporting our conjecture about the missing poles. It indicates that the sheet $+2$ pole is the 2nd Efimov state, as suggested in Fig.~\ref{fig:sheets}. However, we note that this result is obtained in an approximate, model-dependent manner and is less reliable than the exact analytic continuation of the amplitude.


\subsection{Scaling of bound-state energies and vertex functions}

Here, we present additional evidence of the recovery of the discrete scaling symmetry from the three-body integral equations. In agreement with Efimov's prediction, it emerges not only in the unitarity limit but for finite values of $a$ as well. All characteristic features of the three-body states (e.g., energies, trajectories, pole residues) are with good precision described by the same functions of $a$, modulo rescaling of dimensional quantities by appropriate powers of $Q_a$, e.g., $a \to Q_a a$ or $E \to E/Q_a^2$.

\begin{figure}[t]
    \centering
    \includegraphics[width=0.99\textwidth, trim = {0 0 0 0}, clip]{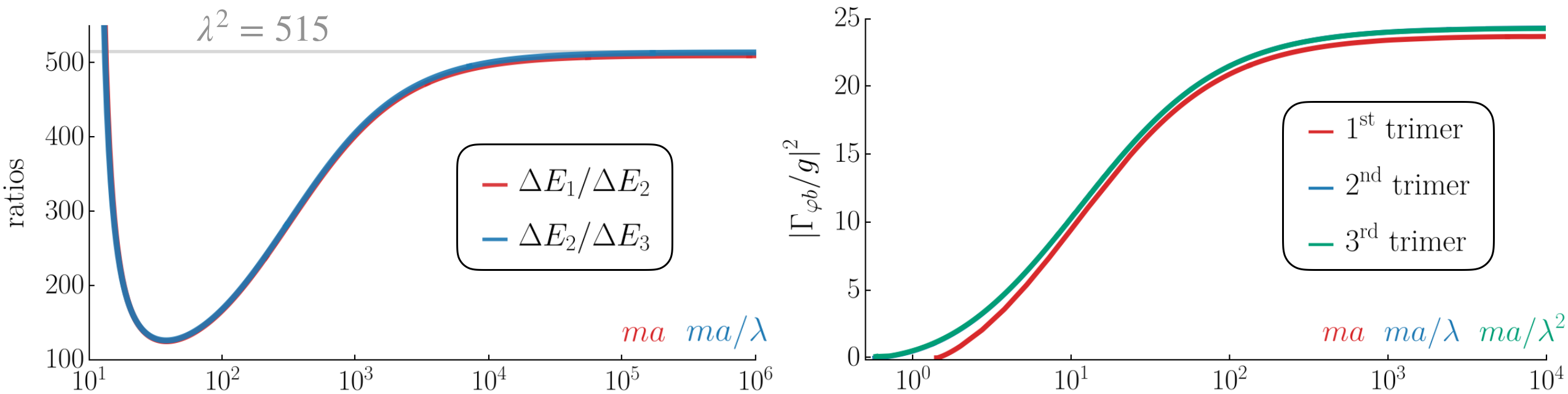}
    \caption{Left panel: ratios of subsequent binding energies as the functions of $ma$. The scattering length for the $\Delta E_2/\Delta E_3$ is rescaled by $\lambda$, as indicated by the label on the horizontal axis. Efimov's ratio, $\lambda^2$, is presented as a gray solid line. Right panel: normalized residues of the first three bound state poles of the $\Mc_{\varphi b}$ amplitude as the functions of $ma$. The scattering lengths for the subsequent residues are rescaled by appropriate powers of $\lambda$, as indicated by the labels on the horizontal axis. }
    \label{fig:9}
\end{figure}

\begin{figure}[t]
    \centering
    \includegraphics[width=0.65\textwidth, trim = {0 0 0 0}, clip]{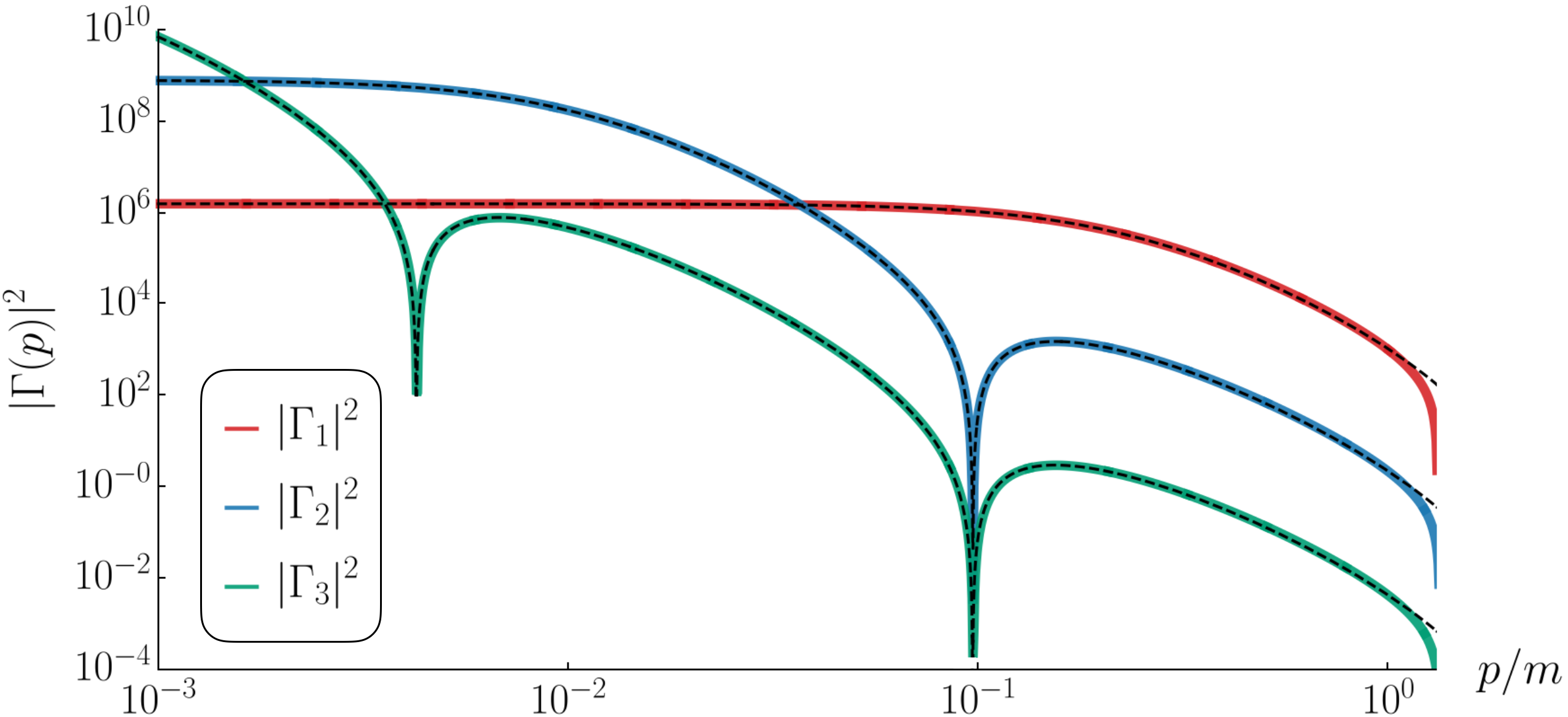}
    \caption{Residues of the $\Mc_3$ amplitude at the first three trimer poles for $ma=-10^{6}$. Dashed black lines represent the analytic prediction of Ref.~\cite{Hansen:2016ync}.}
    \label{fig:10}
\end{figure}

As expected, the bound states in the relativistic system behave like Efimov states when their binding energies become small compared to the scale-invariance breaking quantities like $m$ or the cutoff $k_{\text{max}}$. As shown on the left panel of Fig.~\ref{fig:9}, the ratio of subsequent binding energies approaches the square of Efimov's constant in the unitary limit. Extrapolated values are $0.7\%$ away from $\lambda^2$ for the $\Delta E_1/\Delta E_2$ ratio, and $10^{-3}$\% for the $\Delta E_2/\Delta E_3$ one, which we consider an excellent agreement. For finite values of $ma$, both ratios show similar functional dependence on the scattering length. After we rescale it by a factor of $\lambda$ for the second one, they overlap as shown in Fig.~\ref{fig:10}. Although the energies exhibit dependence on the $\Kc_3$, their ratios are mostly independent of its value, in agreement with the argument made in Sec.~\ref{sec:f_infty} of this supplement.

On the right panel of Fig.~\ref{fig:8}, we show that the analogous property holds for the residues associated with the trimer poles. At positive values of $a$, we compute the $\Mc_{\varphi b}$ amplitude introduced in Eq.~\eqref{eq:LSZ}, and extract the residues from the expansion,
    \beq
    \Mc_{\varphi b} = - \frac{|\Gamma_{\varphi b}|^2}{E^2- E_n^2} + \Oc(E) \, .
    \eeq
Residue $|\Gamma_{\varphi b}|^2$ describes the coupling strength between the trimer and the $\varphi b$ state. We normalize it with the two-body coupling $g^2$ and plot it as a function of $ma$. As we can see, rescaling the characteristic length scales brings all residues close to each other, with almost no difference between the second and the third trimer and a $\Oc(10\%)$ discrepancy between these two and the first one. Despite the relatively small binding energy of the deep bound state, we understand it as a remnant of relativistic effects in the formation of this state.

Finally, in Fig.~\ref{fig:10}, we reproduce numerically the known analytic form for the non-relativistic vertex function, which was derived in Ref.~\cite{Hansen:2016ync} and confirmed numerically in Ref.~\cite{Briceno:2018mlh},
    \begin{align}
    \label{eq:NR-vertex}
    |\Gamma(k)|^2
    =|c||A|^2 \frac{256\pi^{5/2}}{3^{1/4}} \frac{m^2\kappa^2}{k^2(\kappa^2+3k^2/4)} \frac{\sin^2 \Big(s_0 \sinh^{-1} \big(\sqrt{3}k/2\kappa \big) \Big) }{\sinh^2(\pi s_0/2) } \, .
    \end{align}
Here, $\kappa = - \sqrt{|m \Delta E|}$ and $s_0 \approx 1.006$. Normalization constant $|c| = 96.351$, while $A$ is close to $1$ when $ma \to \infty$. The plot in Fig.~\ref{fig:10} was obtained for the first three trimers at $ma=-10^6$, which corresponds to binding energies: $\Delta E_1/m = -2.8 \cdot 10^{-2}$, $\Delta E_2/m = -5.5 \cdot 10^{-5}$, and $\Delta E_3/m = -1.1 \cdot 10^{-7}$. The presented agreement of the non-relativistic result with our finding is another confirmation that the observed states are undoubtedly Efimov in nature. As expected, we can observe a discrepancy between our numerical result and the analytic formula for momenta that can be considered relativistic, $p/m = \Oc(1)$. 

\end{document}